\pdfoutput=1

\newif\ifdraft\draftfalse  
\newif\ifanon\anonfalse    
\newif\iffull\fullfalse   
\newif\ifbackref\backreffalse 
\newif\ifsooner\soonerfalse
\newif\iflater\laterfalse
\newif\ifieee\ieeefalse
\newif\ifcheckpages\checkpagesfalse
\newif\ifdesperateforspace\desperateforspacetrue
\newif\ifcamera\cameratrue
\newif\iflongrefs\longrefsfalse
\makeatletter \@input{texdirectives} \makeatother
\iffull\desperateforspacefalse\fi

\makeatletter\@input{hash}\makeatother

\ifcamera
\RequirePackage{pdf14} 
\fi

\ifieee
\documentclass[10pt, conference, compsocconf, letterpaper, times]{IEEEtran}

\else 

\ifcamera
\documentclass[sigconf]{acmart}
\else
  \ifanon
  \documentclass[sigconf, anonymous, review, nonacm]{acmart}
  \else
  \documentclass[sigconf]{acmart}
  \fi
\fi

\ifanon
\acmConference[Anonymous Submission \#88 to ACM CCS 2018]{ACM Conference on Computer and Communications Security}{Due 8 May 2018}{Toronto, Canada}
\else
\fi

\copyrightyear{2018--2019}
\setcopyright{rightsretained}
\acmConference[CCS '18]{2018 ACM SIGSAC Conference on Computer and Communications Security}{October 15--19, 2018}{Toronto, ON, Canada}

\ifcamera
\acmYear{2018}
\acmBooktitle{2018 ACM SIGSAC Conference on Computer and Communications Security (CCS '18), October 15--19, 2018, Toronto, ON, Canada}
\acmDOI{10.1145/3243734.3243745}
\acmISBN{978-1-4503-5693-0/18/10}
\else
\acmDOI{}
\acmISBN{}
\fi

\ifcamera\else
\makeatletter
\def\@mkbibcitation{}
\makeatother
\fi

\makeatletter
\def\@copyrightpermission{This work is licensed under a \href{https://creativecommons.org/licenses/by/4.0/}{Creative Commons Attribution 4.0 International License}.
Any updated versions will be made available at \url{https://arxiv.org/abs/1802.00588}}
\makeatother

\ifcamera
\makeatletter
\fancypagestyle{firstpagestyle}{%
  \fancyhf{}%
  \fancyhead[L]{}%
  \fancyhead[R]{}%
}
\fancypagestyle{standardpagestyle}{
  \fancyhf{}%
  \fancyhead[L]{}%
  \fancyhead[R]{}%
  \fancyfoot[C]{}%
}
\makeatother
\else
\makeatletter
\fancypagestyle{firstpagestyle}{%
  \fancyhf{}%
  \renewcommand{\headrulewidth}{\z@}%
  \renewcommand{\footrulewidth}{\z@}%
  \fancyhead[L]{\ACM@linecountL}%
  \fancyhead[R]{\ACM@linecountR}%
}
\fancypagestyle{standardpagestyle}{
  \fancyhf{}%
  \renewcommand{\headrulewidth}{\z@}%
  \renewcommand{\footrulewidth}{\z@}%
  \fancyhead[L]{\ACM@linecountL\@headfootfont\shorttitle}%
  \fancyhead[R]{\@headfootfont\@shortauthors\ACM@linecountR}%
  \fancyfoot[C]{\thepage}%
}
\makeatother
\fi

\fi

\usepackage{etex}
\usepackage{mathtools}
\usepackage{xspace}
\usepackage{multirow}
\usepackage{array}
\usepackage{amsmath,amssymb,amsthm}
\usepackage{stmaryrd}
\SetSymbolFont{stmry}{bold}{U}{stmry}{m}{n} 
\usepackage{xcolor}
\usepackage{listings}

\let\ls\lstinline
\usepackage{paralist}
\usepackage{fancyvrb}
\usepackage{latexsym}
\usepackage{bm}
\usepackage{morefloats}
\usepackage[shortlabels,inline]{enumitem}
\usepackage{fancyhdr}
\usepackage[yyyymmdd,hhmmss]{datetime}
\usepackage{graphicx}
\ifieee\usepackage[noadjust]{cite}\fi
\usepackage{coqed}

\usepackage{combelow}

\usepackage{tikz}
\usetikzlibrary{matrix,arrows,decorations.pathmorphing,positioning}

\usepackage{subcaption}

\usepackage{preamble}
\typicallabel{MkKey}

\ifieee\else\fi

\newcommand{\rsc}[0]{\ii{RSC}\xspace}
\newcommand{\rscdc}[0]{\ii{RSC}\ensuremath{^\ii{DC}}\xspace}
\newcommand{\rscdcmd}[0]{\ii{RSC}\ensuremath{^\ii{DC}_\ii{MD}}\xspace}
\newcommand{\rscc}[0]{\ii{RSCC}\xspace}
\newcommand{\bcc}{\ii{BCC}\xspace}
\newcommand{\fcc}{\ii{FCC}\xspace}
\newcommand{\doesprefix}{{\rightsquigarrow}^{*}}

\newcommand{\mypar}[1]{\smallskip\noindent{\em #1\quad}}
\newcommand{\myindentedpar}[1]{\smallskip{\em #1\quad}}

\newcommand{\cmp}[1]{#1\hspace{-0.35em}\downarrow}
\newcommand{\back}[1]{#1\hspace{-0.35em}\uparrow}

\makeatletter
\AtBeginDocument{
  \hypersetup{
    pdftitle = {When Good Components Go Bad: Formally Secure Compilation Despite Dynamic Compromise},
    pdfauthor = {\@author}
  }
}
\makeatother

\ifieee
\title{\bf{\Huge When Good Components Go Bad}\\[0.5ex]
\LARGE Formally Secure Compilation Despite Dynamic Compromise\ifanon\ifieee\vspace{-1.5em}\fi\fi}
\else
\title{When Good Components Go Bad}
\subtitle{\ifcamera\else\bf\huge\fi
  Formally Secure Compilation Despite Dynamic Compromise}
\fi

\ifcamera
\author{Carmine Abate}
\affiliation{
  \institution{Inria Paris}
  \city{Paris}\country{France}}

\author{Arthur Azevedo de Amorim}
\affiliation{
  \institution{CMU}
  \city{Pittsburgh}\country{USA}}

\author{Roberto Blanco}
\affiliation{
  \institution{Inria Paris}
  \city{Paris}\country{France}}

\author{Ana Nora Evans}
\affiliation{
  \institution{Inria Paris}
  \city{Charlottesville}\country{USA}}

\author{Guglielmo Fachini}
\affiliation{
  \institution{Nozomi Networks}
    \city{Mendrisio}\country{Switzerland}}

\author{C\u{a}t\u{a}lin Hri\cb{t}cu}
\affiliation{
  \institution{Inria Paris}
  \city{Paris}\country{France}}

\author{Th\'eo Laurent}
\affiliation{
  \institution{Inria Paris and ENS Paris}
  \city{Paris}\country{France}}

\author{Benjamin C. Pierce}
\affiliation{
  \institution{Univ. of Pennsylvania}
  \city{Philadelphia}\country{USA}}

\author{Marco Stronati}
\affiliation{
  \institution{Nomadic Labs}
  \city{Paris}\country{France}}

\author{J\'er\'emy Thibault}
\affiliation{
  \institution{Inria Paris}
  \city{Paris}\country{France}}

\author{Andrew Tolmach}
\affiliation{
  \institution{Portland State Univ.}
  \city{Portland}\country{USA}}

\else
\author{
\ifanon\else
\ifieee\else\large\fi
  Carmine Abate\textsuperscript{1} \quad
  Arthur Azevedo de Amorim\textsuperscript{2} \quad
  Roberto Blanco\textsuperscript{1} \quad
  Ana Nora Evans\textsuperscript{3} \quad
  Guglielmo Fachini\textsuperscript{4} \\
  C\u{a}t\u{a}lin Hri\cb{t}cu\textsuperscript{1} \quad
  Th\'eo Laurent\textsuperscript{1} \quad
  Benjamin C. Pierce\textsuperscript{5} \quad
  Marco Stronati\textsuperscript{6} \quad
  J\'er\'emy Thibault\textsuperscript{1} \quad
  Andrew Tolmach\textsuperscript{7}\\[2ex]
  \textsuperscript{1}Inria Paris\quad
  \textsuperscript{2}Carnegie Mellon University\quad
  \textsuperscript{3}University of Virginia\quad
  \textsuperscript{4}Nozomi Networks\\
  \textsuperscript{5}University of Pennsylvania\quad
  \textsuperscript{6}Nomadic Labs\quad
  \textsuperscript{7}Portland State University\ifieee\\[-1ex]\else\\[1ex]\fi
\fi
}
\fi
\makeatletter
\renewcommand{\@shortauthors}{Abate~\ETAL}
\makeatother

\ifcamera

\fi


\begin{document}

\begin{abstract}
We propose a new formal criterion for evaluating {\em secure compilation}
schemes for unsafe languages, expressing end-to-end security guarantees for
software components that may become compromised after encountering {\em
undefined behavior}---for example, by accessing an array out of bounds.

Our criterion is the first to model {\em dynamic} compromise in a system of
mutually distrustful components with clearly specified privileges. It
articulates how each component should be protected from all the others---in
particular, from components that have encountered undefined behavior and
become compromised.
Each component receives secure compilation guarantees---in particular, its
internal invariants are protected from compromised components---up
to the point when this component itself becomes compromised, after
which we assume an attacker can take complete control and use this
component's privileges to attack other components.
More precisely, a secure
compilation chain must ensure that a dynamically compromised component
cannot break the safety properties of the system at the target level any
more than an arbitrary attacker-controlled component (with the same
interface and privileges, but without undefined behaviors) already could at
the source level.

To illustrate the model, we construct a secure compilation chain for a
small unsafe language with buffers, procedures, and components, targeting a
simple abstract machine with built-in compartmentalization.  We give a
machine-checked proof in Coq that this
compiler satisfies our secure compilation criterion.  Finally, we show that
the protection guarantees offered by the compartmentalized abstract machine
can be achieved at the machine-code level using either software fault
isolation or a tag-based reference monitor.

\ifsooner
\ch{exhibit: I though we settled on ``encounter'' for this}\bcp{I'd forgotten
that we'd settled on it, though I did think of "encounter" as an
alternative phrasing here.  The reason it feels less good to me, now, is
that it suggests that this component is "encountering" someone {\em else}'s
undefined behavior.  I.e., "encounter" isn't normally used for something
that you do to yourself---it means "meet" literally.}\ch{I'm fine with
  any of these as long as we use it consistently. Would need to look
  at more than a single context in which one fits better than the other.}
\fi
\end{abstract}

\maketitle

\section{Introduction}
\label{sec:intro}

\hyphenation{Web-Assembly Comp-Cert}

{\em Compartmentalization} offers a strong, practical defense against a range of
devastating low-level attacks, such as control-flow hijacks exploiting
buffer overflows and other vulnerabilities in C, C++, and other unsafe
languages~\cite{GudkaWACDLMNR15, WatsonWNMACDDGL15, wedge_nsdi2008}.
%
Widely deployed compartmentalization technologies include process-level privilege
separation~\cite{Kilpatrick03, GudkaWACDLMNR15, wedge_nsdi2008} (used
in OpenSSH~\cite{ProvosFH03} and for sandboxing plugins and tabs
in web browsers~\cite{ReisG09}), software fault
isolation~\cite{sfi_sosp1993, Tan17} (\EG Google Native
Client~\cite{YeeSDCMOONF10}),
WebAssembly modules~\cite{HaasRSTHGWZB17} in modern web browsers,
and hardware enclaves (\EG Intel SGX~\cite{sgx});
many more are on the drawing
boards~\cite{micropolicies2015,WatsonWNMACDDGL15,cheri_asplos2015,
  SkorstengaardDB18}.
These mechanisms offer an attractive base for building more secure
compilation chains that mitigate low-level
attacks~\cite{JuglaretHAPST15, PatrignaniDP16, GudkaWACDLMNR15, TsampasEPDGP,
vanGinkelSMP16, StrydonckDP18, GollamudiF18\ifdesperateforspace\else\fi}.
In particular, compartmentalization can be applied
%
in unsafe low-level languages to structure large, performance-critical
applications into mutually distrustful components that have clearly specified
privileges and interact via well-defined interfaces.

Intuitively, protecting each component from all the others should bring
strong security benefits, since
a vulnerability in one component need not
compromise the security of the whole application.
Each component will be protected from all other components for as long as it
remains ``good.''
If, at some point, it encounters an internal vulnerability such as a buffer overflow, then, from
this point on, it is assumed to be
compromised and under the control of the attacker, potentially causing it to
attack the remaining uncompromised components.  The main goal of this
paper is to formalize this dynamic-compromise intuition and precisely
characterize what it means for a compilation chain to be secure in this
setting.

\iflater
\ch{Andrew asked if we can add some more upfront intuition and maybe an example}
\fi

We want a characterization that supports {\em source-level
  security reasoning}, allowing programmers to reason
about the security properties of their code without knowing anything
about the complex internals of the compilation chain
(compiler, linker, loader, runtime system, system software, etc).
What makes this particularly challenging for C and C++ programs is that
they may encounter {\em undefined behaviors}---situations that have no
source-level meaning whatsoever.
%
Compilers are allowed to assume
that undefined behaviors never occur in programs, and they
aggressively exploit this assumption
to produce the fastest possible code for well-defined
programs, in particular by avoiding the insertion of run-time checks.
For example, memory safety
violations~\cite{Szekeres2013, memory-safety} (\EG accessing an array out of
bounds, or using a pointer after its memory region has been freed) and type
safety violations~\cite{DuckY18, HallerJPPGBK16} (\EG invalid unchecked
casts)---cause real C compilers to produce code that behaves arbitrarily,
often leading to exploitable
vulnerabilities~\cite{Szekeres2013, Heartbleed}.

Of course, not every undefined behavior is necessarily exploitable.
However, for the sake of strong security guarantees, we make a worst-case assumption
that {\em any} undefined behavior encountered within a component can lead to its
compromise.
%
Indeed, in the remainder of the paper we equate the notions of ``encountering
undefined behavior'' and ``becoming compromised.''

While the dangers of memory safety and casting violations are widely understood,
the C and C++ standards~\cite{c11} call out large numbers of undefined
behaviors~\cite{HathhornER15, Krebbers15} that are less familiar,
even to experienced C/C++
developers~\cite{Lattner11, WangZKS13\iffull, WangCCJZK12\fi}.
To minimize programmer confusion and lower the risk of introducing
security vulnerabilities,
real compilers generally give sane and predictable semantics to
{\em some} of these behaviors.
For example, signed integer overflow is officially an undefined behavior in standard C,
but many compilers (at least with certain flags set) guarantee that the
result will be calculated using wraparound arithmetic.
Thus, for purposes of defining secure compilation, the set of undefined
behaviors is effectively defined by the compiler at hand rather than by the standard.
\iflater
\apt{This para. may be important, but it feels like it should be a footnote.}
\ch{Looks good to me as a paragraph. It redefines a standard term,
  so the paper can't be properly understood without it; while in general
  I expect footnotes to be skippable.}
\fi

\ifsooner
\bcp{The transition between the previous para and the next is clunky.}
\fi

The purpose of a compartmentalizing compilation chain is to ensure that the
arbitrary, potentially malicious, effects of undefined behavior are limited
to the component in which it occurs.
For a start, it should restrict the {\em spatial} scope of a compromise to
the component that encounters undefined behavior.
Such compromised components can only influence
other components via controlled interactions respecting their interfaces and
the other abstractions of the source language
(\EG the stack discipline on calls and returns).
%
Moreover, to model dynamic compromise and give each component full
guarantees as long as it has not yet encountered undefined behavior, the {\em
  temporal} scope of compromise must also be restricted.
In particular, compiler optimizations should never cause the effects of
undefined behavior to show up before earlier ``observable events'' such as
system calls.
Unlike the spatial restriction, which requires some form of run-time enforcement
in software or hardware, the temporal restriction can be enforced just by
foregoing certain aggressive optimizations.
For example, the temporal restriction (but not the spatial one)
is already enforced by the CompCert C
compiler~\cite{Leroy09, Regehr10},
providing a significantly cleaner model of undefined behavior than other C
compilers~\cite{Regehr10}.


We want a characterization that is {\em formal}---that
brings mathematical precision to the security guarantees
and attacker model of compartmentalizing compilation.
This can serve both as a clear specification for verified secure
compilation chains and as useful guidance for unverified ones.
Moreover, we want the characterization to provide
source-level
reasoning principles that can be used to assess the security of
compartmentalized applications.
%
To make this feasible in practice, the {\em amount} of source code
to be verified or audited has to be relatively small.
%
So, while we can require developers to carefully analyze the privileges of
each component and the correctness of some very small pieces of
security-critical code, we cannot expect them to establish the full
correctness---or even absence of undefined behavior---for most of their
components.

Our secure compilation criterion improves on the state of the art in three
important respects.
First, our criterion applies to {\em compartmentalized} programs, while
most existing formal criteria for secure compilation are phrased in terms of
protecting a single trusted program from an untrusted context
~\cite{AbadiP13, AbadiP12, PatrignaniASJCP15, FournetSCDSL13, Abadi99,
AgtenSJP12, Agten0P15,
  \ifdesperateforspace\else LarmuseauPC15,\fi AbateBGHPT19}.
Second, unlike some recent criteria
that do consider modular protection~\cite{DevriesePPK17,
  PatrignaniDP16}, our criterion applies to {\em unsafe} source languages
with undefined behaviors.
And third, it considers a {\em dynamic} compromise model---a critical
advance over the recent proposal of Juglaret~\ETAL~\cite{JuglaretHAEP16},
which does consider components written in unsafe languages,
but which is limited to a {static} compromise model.
This is a serious limitation: components whose code contains any
vulnerability that might potentially manifest itself as undefined behavior
are given no guarantees whatsoever, irrespective of whether an attacker
actually exploits these vulnerabilities.
Moreover, vulnerable components lose all guarantees from the start of
the execution---possibly long before any actual compromise.
Experience shows that large enough C or C++ codebases essentially
always contain vulnerabilities~\cite{Szekeres2013}.  Thus, although static
compromise models may be appropriate for safe languages, they are \iffull
generally \fi not useful \iffull in practice \fi for unsafe low-level languages.
%

As we will see in \autoref{sec:related},
the limitation to static compromise scenarios seems inescapable for
previous techniques, which are all based on the formal criterion of
{\em full abstraction}~\cite{Abadi99}.
%
%
To support {dynamic} compromise scenarios, we take an unconventional
approach, dropping full abstraction and instead phrasing our criterion in
terms of preserving safety properties~\cite{LamportS84}
in adversarial contexts~\cite{AbateBGHPT19},
%
%
where, formally, safety properties are predicates over execution traces
that are informative enough to detect the compromise of components and
to allow the execution to be ``rewound'' along the same trace.
%
%
Moving away from full abstraction also makes our criterion easier
to achieve efficiently in practice and to prove at scale.
\ch{Temporarily removed this part, see comments below:
Finally, we expect our criterion to scale naturally from trace properties to
hyperproperties such as confidentiality~\cite{AbateBGHPT19}
(see \autoref{sec:related} and \autoref{sec:conclusion}).

\ch{Can't this work also be extended to Robust 2-relational Safety
  Preservation (R2rSP)? (In the absence of side-channels) Because if
  it could then it would already be stronger than full abstraction.
  Question: what would that buy us? Do real developers really care
  about hiding code?}

\ch{Warning: the interaction between dynamic compromise and
  hyperproperties is more subtle than I originally though. Might want
  to reduce useless speculation here.}}





\ifieee
\paragraph{Contributions}
\else
\smallskip\noindent
{\bf Contributions.}\quad
%
Our first contribution is {\em Robustly Safe Compartmentalizing
  Compilation (\rscc)},
  a new secure compilation criterion
  articulating strong end-to-end security guarantees for components written in
  unsafe languages with
  undefined behavior. This
  criterion is the first to support {\em dynamic compromise} in a
  system of {\em mutually distrustful components} with clearly
    specified privileges.
%
%
%
  We start by illustrating the intuition, informal attacker model,
  and source-level \iffull security \fi
  reasoning \iffull principles \fi
  behind \rscc using a simple example application (\autoref{sec:rcc-informal}).


  Our second contribution is a formal presentation of \rscc.
%
  We start from {\em Robustly Safe Compilation} (\rsc,
  \autoref{sec:rsc}), a simple security criterion recently introduced
  by Abate~\ETAL\cite{AbateBGHPT19}, and
  extend this first to dynamic compromise (\rscdc, \autoref{sec:rsc-dc}),
  then mutually distrustful components (\rscdcmd, \autoref{sec:rsc-dc-md}),
  and finally to the full definition of \rscc (\autoref{sec:rcc-formal}).
  We also give an effective and generic proof technique for \rscc
  (\autoref{sec:proof-technique}):
%
%
  We start with a target-level execution and explain any finite sequence
  of calls and returns in terms of the source language by constructing
  a whole source program that produces this prefix.
%
%
  Our novel proof architecture manages to achieve this using only (mostly
  standard) simulation proofs, which is much simpler and more scalable than
  previous proofs in this space~\cite{JuglaretHAEP16}.
  One particularly important advantage is that it allows us to reuse a
  whole-program compiler correctness result \`a la CompCert~\cite{Leroy09}
  as a black box, avoiding the need to prove any other simulations between
  the source and target languages.
%
%

  Our third contribution is a proof-of-concept secure
  compilation chain (\autoref{sec:compiler})
%
%
  for a simple unsafe sequential language
  featuring buffers, procedures, components, and a
  CompCert-like block-based memory model~\cite{LeroyB08} (\autoref{sec:source}).
  Our entire compilation chain is implemented
  in the Coq proof assistant.
  The first step compiles our source language to a simple low-level
  abstract machine with built-in compartmentalization (\autoref{sec:intermediate}).
  We use the proof technique from \autoref{sec:proof-technique} to construct
  machine-checked proofs in Coq showing that this compiler satisfies
  \rscc~(\autoref{sec:proving-rscdcmd-instance}).
%
  Finally, we describe two back ends for our compiler, showing that the
  protection guarantees of the compartmentalized abstract machine can be
  achieved at the lowest level using either software fault isolation (SFI,
  \autoref{sec:sfi}) or a tag-based reference monitor
  (\autoref{sec:micro-policy}).
%
%
  The tag-based back end, in particular, is novel, using linear return capabilities
  to enforce a cross-component call/return discipline.
  Neither back end has yet been formally verified, but
  we have used property-based testing to gain confidence that the SFI
  back end satisfies \rscdcmd.
\iflater\ch{Any chance we can test \rscdcmd for both backends?}\fi{}


These contributions lay a solid foundation for future secure
compilation chains that could bring sound and practical
compartmentalization to C, C++, and other unsafe low-level languages.
We address three fundamental questions:
{\em (1) What is the desired secure compilation criterion and to what
  attacker model and source-level security reasoning
  principles does it correspond?}  Answer: We propose the \rscc criterion
from \autoref{sec:rcc-informal}-\autoref{sec:formal}.
{\em (2) How can we effectively enforce secure compilation?}
  Answer: Various mechanisms are possible; the simple compilation
  chain from \autoref{sec:compiler} illustrates how either software fault isolation
  or tagged-based reference monitoring can enforce \rscc.
{\em (3) How can we achieve high assurance that the resulting
     compilation chain is indeed secure?}
Answer: We show that formal
    verification (
    \autoref{sec:proving-rscdcmd-instance})
     and property-based testing (\autoref{sec:sfi})\iflater\ch{new
       subsection if doing both backends}\fi{}
     can be successfully used together for this in a proof assistant like Coq.

We close with \iffull a discussion of \fi related
(\autoref{sec:related}) and future~(\autoref{sec:conclusion}) work.
The appendix presents omitted \iffull technical \fi details.
Our Coq development is available
\ifanon anonymously\fi at
\ifanon\url{https://github.com/WhenGoodComponentsGoBad/Coq/tree/\hash}\else
{\footnotesize \tt \href{https://github.com/secure-compilation/when-good-components-go-bad/}{https://github.com/secure-compilation/when-good-components-go-bad/}}\fi
%
%
\ifanon
, together with a summary of improvements to the paper in
response to helpful reviewer feedback on a previous IEEE S\&P submission%
\fi

A previous version of this work has been published in conference
proceedings~\cite{AbateABEFHLPST18}.
The current version has been extended to include the details of the proofs, and,
compared to the conference version, more proofs are now fully mechanized in
Coq, and our proof architecture has also been significantly simplified.
\ch{Florian's thing}

\iflater
\ch{Do we want to also send them some paper proofs?  I have something
  for \autoref{sec:rcc-formal}, but I'm already sketching the main
  idea in the text.}
\bcp{Probably not enough time?  But if there are any proofs that we don't
  have time to formalize, it might be a good idea to send a paper sketch.}
\fi

\ifcheckpages\clearpage\fi
\section{\rscc By Example}
\label{sec:rcc-informal}


\iflater
\apt{This example is adequate, but the paper would be greatly strengthened if we
  we could provide a ``killer example'' whose security relevance and importance are obvious.}
\fi

\ifsooner\ch{Too many fonts in this section.}\fi

We begin by an overview of compartmentalizing compilation chains, our attacker
model, and how viewing this model as a dynamic compromise game leads to
intuitive principles for security analysis.



%
We need not be very precise, here, about the details of the source language;
we just assume that it is equipped with some compartmentalization
facility~\cite{GudkaWACDLMNR15, VasilakisKRDDS18} that allows programmers to break up
security-critical applications into mutually distrustful {\em components}
that have clearly specified {\em privileges} and can only interact via
well-defined {\em interfaces}.
In fact we assume that the interface of each component gives a precise
description of its privilege.\ifsooner\bcp{Is this an assumption, or would it be
  better to say that we {\em define} the privilege of a component as the set
of external facilities it is allowed to invoke?
\ch{Not sure about this, all we assume is that what we call the interface
  also captures the privileges, not that it can't capture other things
  like types  that are less ``privilege like''.\bcp{What's not
    privilege-like about types? :-)} This might
  not come across with the current phrasing either though.}
(Also, a reader new to our
work may wonder how this corresponds to the usual intuition of
privilege---i.e., ``privilege''  $\sim$ ``the power to see private information
coming from outside or cause things to happen outside the system'').  It
sounds like we're saying, for example, that the power to write something on
the surface of the disk is not an example of privilege.}\fi{}
The notions of component and interface that we use for defining the secure
compilation criteria in \autoref{sec:formal} are quite generic:
interfaces can include any
requirements that can be enforced on components, including type
signatures, lists of allowed system calls, or more detailed access-control
specifications describing legal parameters to cross-component
calls (\EG ACLs for operations on files).
We assume that the division of an application into components and the
interfaces of those components are statically determined and fixed\iffull
throughout execution\fi.\ifsooner\bcp{We could put a footnote here remarking that we
  discuss the possibility of extending to dynamic component creation in the
  future work section.}\ch{No strong feelings, we can add but it's
  a bit of detour}\fi{}
For the illustrative language of \autoref{sec:compiler}, we will use a
simple setup in which components don't directly share
state, interfaces just list the procedures that each component provides
and those that it expects to be present in its context,
and the only thing one component can do to another one is
to call procedures allowed by their interfaces.

The goal of a compartmentalizing compilation chain is to ensure that
components interact according to their interfaces {even} in the presence of
undefined behavior.  Our secure compilation criterion\ifsooner\bcp{as discussed above,
  I find phrases
  like ``security criterion'' confusing because of the two possible meanings
  of ``security.''  I think it would be best to always say ``secure
  compilation criterion'' to avoid this confusion}\bcp{I'm still
  surprised by the sudden switch to ``criteria'' here (and one instance
  above).  I guess it refers to the intermediate definitions in section 4,
  but I thought we were not talking about them in this section: every time I
  read this I start wondering why we've switched from one top-level property
  to (apparently) more than one.}\ch{Fair points, fixed above}\fi
does not fix a
specific mechanism for achieving this\iffull guarantee\fi:
responsibility can be divided among the different parts of the
compilation chain, such as the compiler, linker, loader, runtime system, system
software, and hardware.
In \autoref{sec:implementing} we study a compilation chain with two
alternative back ends---one using software fault isolation
and one using tag-based reference monitoring for compartmentalization.
What a compromised component {\em can} still do in this model is to use its
access to other components, as allowed by its interface, to either trick them into
misusing their own privileges (\IE confused deputy attacks) or even compromise
them as well (\EG by sending them malformed inputs that trigger
control-hijacking attacks exploiting undefined behaviors in their code).

\ifsooner
\ch{This doesn't seem super general, but I also can't attribute it to
  our instance until we actually implement it there. Might attribute
  it to the example for now?}%
\bcp{Sadly, my first thought on re-reading this part was ``only in the
  examples?  what about in the formalization?''}%
\ch{Without an extension of the Coq instance, all I can think of
  is to just be vague and drop ``In the examples below,''}%
\bcp{Better.}
\fi
We model input and output as interaction with a designated {\em
  environment} component {\tt E} that is given an interface but no implementation.
When invoked, environment functions are assumed to immediately
return a non-deterministically chosen value~\cite{Leroy09}.
In terms of security, the environment is thus the initial source of arbitrary,
possibly malformed, inputs that can exploit buffer overflows and other
vulnerabilities to compromise other components.\ifsooner\bcp{Wearing my
  ``theoretical
  security researcher'' hat, I keep wanting to ask ``What observations of
  the whole system do we use to judge whether secrets are being exposed on
  public channels, missiles are being launched based on low-integrity
  inputs, or whatever?''}\fi

As we argued in the introduction, it is often unrealistic to assume that we know in
advance which components will be compromised and which ones will not.
This motivates our model of {\em dynamic compromise}, in which
each component receives secure compilation guarantees until it becomes
compromised by encountering
an undefined behavior, causing it to start attacking the remaining
uncompromised components.
In contrast to earlier static-compromise models~\cite{JuglaretHAEP16}, a
component only loses guarantees in our model after an attacker discovers and
manages to exploit a vulnerability, by sending it inputs that lead to an
undefined behavior.
The mere existence of vulnerabilities---undefined behaviors that can be reached
after some sequence of inputs---is not enough for the component to be considered
compromised.

This model allows developers to reason informally about various
compromise scenarios and their impact on the security
of the
whole application~\cite{GudkaWACDLMNR15}\iffull, relying on informal reasoning of
the form: ``If component A gets taken over then it can compromise
component B, which can then compromise C, which means
that {\em this} bad thing can happen (while {\em that} other bad thing
still cannot), whereas if these other components get taken over, then
{\em this} other bad thing might happen...''\apt{an example would be better
  than this blah-blah-blah description, evocative though it is.}\ch{The
  example is coming up below, but it takes more work to set up. Wouldn't
  be against making this more concrete if we find a way to change A, B, C
  and the bad things into obvious to understand instances.}
\else
.
\fi
If the consequences of some plausible compromise seem too serious, developers
can further reduce or separate privilege by narrowing interfaces or splitting
components, or they can make components more defensive by
\iffull dynamically \fi validating their inputs.

\iflater
\nora{At first read, late at night, this seems quite hard to follow.
  It might be clearer to start with what the entire application should do,
  read something, process in some way and print it out. Then proceed with
  explaining what each component should do, and which functions uses for that.}
\ch{I never had any high-level goals for this example application
  beyond illustrating various things. If others think that explaining
  that this application ``reads something, processes it in some
  unspecified way and then writes out the result'' would actually
  clarify things, sure we can say it. Alternatively, we could think of
  a realistic practical scenario in which this kind of processing happens and
  try to make things more specific.}
\fi

As a first running example, consider the idealized
application in \autoref{fig:example-components}.
It defines three components (\ls|C$_0$|, \ls|C$_1$|, and
\ls|C$_2$|\iflater\bcp{It would be easier to understand the example if $C_0$ were
  the one with the main function and $C_2$ were the one with the validation
  function.  I.e., I would rename 1 to 0, 2 to 1, and 0 to 2.}\ch{Would this
  be good for Figure 2 too?}\bcp{Less good. But see my other concern... :-)}\fi) that
interact with the environment \ls$E$ via input (\ls$E.read$) and
output (\ls$E.write$) operations.
Component \ls|C$_1$| defines a \ls$main()$ procedure, which first
invokes \ls|C$_2$.init()| and then reads a request \ls$x$ from the environment (\EG
coming from some remote client), parses it by calling an internal
procedure to obtain \ls$y$, and then invokes \ls|C$_2$.process(x,y)|.
This, in turn, calls \ls|C$_2$.prepare()| and \ls|C$_2$.handle(y)|,
obtaining some \ls|data| that it validates using \ls|C$_0$.valid| and, if this
succeeds, writes
\ls|data| together with the original request \ls$x$ to the
environment\iffull (\EG to be stored on disk)\fi.
%
%
%
\iflater
\ch{Originally,
  having the validation be done by the environment is quite odd, but
  it needed to be external to \ls|C$_2$| to appear in the trace.
  Maybe it's a sign for the future that we should be more permissive
  with components causing events on their own.
  For now introduced a \ls|C$_0$| component instead.}
\fi

\begin{figure}[t]
\begin{lstlisting}
component C$_0$ {
  export valid;
  valid(data) { ... }
}
component C$_1$ {
  import E.read, C$_2$.init, C$_2$.process;
  main() {
    C$_2$.init();
    x := E.read();
    y := C$_1$.parse(x);    //($\color{dkgray}V_1$) can yield Undef for some x
    C$_2$.process(x,y);
  }
  parse(x) { ... }
}
component C$_2$ {
  import E.write, C$_0$.valid;
  export init, process;
  init() { ... }
  process(x,y) {
    C$_2$.prepare();        //($\color{dkgray}V_2$) can yield Undef if not initialized
    data := C$_2$.handle(y);//($\color{dkgray}V_3$) can yield Undef for some y
    if C$_0$.valid(data) then E.write(<data,x>)
  }
  prepare() { ... }
  handle(y) { ... }
}
\end{lstlisting}
\vspace{-1.5em}
\caption{Pseudocode of compartmentalized application}
\label{fig:example-components}
\vspace{-1em}
\end{figure}

Suppose we would like to establish two properties:
\begin{itemize}
\item[($S_1$)] any call \ls|E.write(<data,x>)| happens as a response to a
previous \ls|E.read()| call by \ls|C$_1$| obtaining the request \ls$x$; and
  \item[($S_2$)] the application only writes valid data
(i.e., data for which \ls|C$_0$.valid| returns true).
\end{itemize}
These can be shown to hold of executions that do not encounter undefined behavior
simply by analyzing the control flow. But what if undefined behavior does occur?
Suppose that we can rule out this possibility---by auditing, testing,
or formal verification---for some parts of the code,
but we are unsure about three subroutines:
\begin{itemize}
\item[($V_1$)] \ls|C$_1$.parse(x)| performs complex array computations,
and we do not
  know if it is immune to buffer overflows for all \ls|x|;
\item[($V_2$)] \ls|C$_2$.prepare()| is intended to be called only if
  \ls|C$_2$.init()| has been called beforehand to set up a shared data
  structure; otherwise, it might dereference
  an undefined pointer;
\item[($V_3$)] \ls|C$_2$.handle(y)| might cause integer overflow on some inputs.
\end{itemize}

If the attacker finds an input that causes the
undefined behavior in $V_1$ to occur, then \ls|C$_1$| can get compromised
and call \ls|C$_2$.process(x,y)| with values of \ls|x| that it hasn't received
from the environment, thus invalidating $S_1$.  Nevertheless, if no other
undefined behavior is encountered during the execution, this attack cannot have
any effect on the code run by $C_2$, so $S_2$ remains true.

Now consider the possible undefined behavior from $V_2$. If \ls|C$_1$| is not
compromised, this undefined behavior cannot occur, since \ls|C$_2$.init()| will
be called before \ls|C$_2$.prepare()|. Moreover, this undefined behavior cannot
occur even if \ls|C$_1$| \emph{is} compromised by
\iffull encountering \fi the undefined
behavior in $V_1$, because that can only occur \emph{after} \ls|C$_2$.init()|
has been called.  Hence $V_1$ and $V_2$ together are no worse than $V_1$ alone,
and property $S_2$ remains true.  Inferring this
\iffull property \fi crucially depends on
our model of dynamic compromise, in which \ls|C$_1$| can be treated as honest
and gets \iffull full \fi guarantees until it encounters undefined behavior.
%
If instead we were 
only allowed to reason about \ls|C$_1$|'s ability to do damage based on
its \emph{interface}, as would happen in a model of static
compromise~\cite{JuglaretHAEP16}, we wouldn't be able to conclude that
\ls|C$_2$| cannot be compromised:
an arbitrary component with the same interface as \ls|C$_1$|
could indeed compromise \ls|C$_2$| by calling \ls|C$_2$.process|
before \ls|C$_2$.init|.
Finally, if execution encounters undefined behavior in $V_3$,
then \ls|C$_2$| can get compromised
irrespective of whether \ls|C$_1$| is compromised beforehand,
invalidating both $S_1$ and $S_2$.

\begin{figure}[t]
\begin{lstlisting}
component C$_0$ {
  import E.read, E.write, C$_2$.init, C$_1$.parse, C$_2$.process;
  main() {
    C$_2$.init();
    x := E.read();
    y := C$_1$.parse(x);
    data := C$_2$.process(y);
    if C$_0$.valid(data) then E.write(<data,x>)
  }
  valid(data) { ... }
}
component C$_1$ {
  export parse;
  parse(x) { ... }        //($\color{dkgray}V_1$) can yield Undef for some x
}
component C$_2$ {
  export init, process;
  init() { ... }
  process(y) {
    C$_2$.prepare();        //($\color{dkgray}V_2$) can yield Undef if not initialized
    return C$_2$.handle(y); //($\color{dkgray}V_3$) can yield Undef for some y
  }
  prepare() { ... }
  handle(y) { ... }
}
\end{lstlisting}
\vspace{-0.8em}
\caption{More secure refactoring of the application}
\label{fig:example-components-secure}
\vspace{-1em}
\end{figure}

Though we have not yet made it formal, this security analysis already
identifies \ls|C$_2$| as a single point of failure for both desired properties
of our system.
%
This suggests several ways the program could be improved:
The code in \ls|C$_2$.handle| could be hardened to reduce its chances
of encountering undefined behavior, e.g. by doing better input validation.
%
Or \ls|C$_1$| could validate the values it sends to
\ls|C$_2$.process|, so that an attacker would have to compromise
both \ls|C$_1$| and \ls|C$_2$| to break the validity of writes.
%
To ensure the correspondence of reads and writes despite the
compromise of \ls|C$_1$|, we could make \ls|C$_2$| read the
request values directly from \ls$E$, instead of via \ls|C$_1$|.


\ifsooner
\bcp{I think the reader is supposed to understand that the two variants have ``the
  same component structure'' with just some small bits of code moved around,
but I find this rather unintuitive.  In particular, $C_0$ used to
be just a data-validation module, but now it's become the top level of the
whole system.}
\fi

To achieve the best security though, we can refactor so that the read and write privileges
are isolated in \ls|C$_0$|, which performs no complex data
processing and thus is a lot less likely to be compromised by
undefined behavior (\autoref{fig:example-components-secure}).
%
In this variant, \ls|C$_0$| reads a request, calls \ls|C$_1$.parse| on this
request, passes the result to \ls|C$_2$.process|,
validates the data \ls|C$_2$| returns and then writes it out.
This way both our desired properties hold even if
both \ls|C$_1$| and \ls|C$_2$| are compromised, since now the core application logic and
privileges have been completely separated from the dangerous data
processing operations that could cause vulnerabilities.

\smallskip

Let's begin making all this a bit more formal.  The first step is to make
the security goals of our example application more precise.
We do this in terms of execution {\em traces} that are built from
{\em events} such as cross-component calls and returns.
%
%
The two intuitive properties from our example can be phrased in
terms of traces as follows:
If \ls|E.write(<data,x>)| appears in an execution trace\iffull of the
  program \fi, then \iffull it must be that \fi
\begin{itemize}
\item[$(S_1)$] \ls|E.read| was called previously and returned
\ls|x|, and
\item[$(S_2)$] \ls|C$_0$.valid(data)| was called previously and
returned \ls|true|.
\end{itemize}
The refactored application in \autoref{fig:example-components-secure}
achieves both properties despite the compromise of
both \ls|C$_1$| via $V_1$ and \ls|C$_2$| via $V_3$,
but, for the first variant in
\autoref{fig:example-components} the properties need to be weakened as follows:
If \ls|E.write(<data,x>)| appears in an execution trace then
\begin{itemize}
\item[$(W_1)$] \ls|E.read| previously returned \ls|x| \iffull\\\fi
  {\em or} \ls|E.read| previously
  returned an \ls|x'| that can cause undefined behavior in
  \ls|C$_1$.parse(x')| \iffull\\\fi
  {\em or} \ls|C$_2$.process(x,y)| was called
  previously with a \ls|y| that can cause undefined behavior in
  \ls|C$_2$.handle(y)|, and
\item[$(W_2)$] \ls|C$_0$.valid(data)| was called previously and returned
\ls|true| \iffull\\ \fi
  {\em or} \ls|C$_2$.process(x,y)| was called previously with a \ls|y|
  that can cause undefined behavior in \ls|C$_2$.handle(y)|.
\end{itemize}
While these properties are significantly weaker (and harder to understand),
they are still not trivial; in particular, they still tell us something useful
under the assumption that the attacker has not actually discovered how to
compromise \ls|C$_1$| or \ls|C$_2$|.

Properties $S_1$, $S_2$, $W_1$, \iffull and \fi $W_2$ are
all {\em safety properties}
\cite{LamportS84}---inspired, in this case, by the sorts of ``correspondence
assertions'' used to specify authenticity in security protocols
\cite{WooL93, GordonJ04\ifdesperateforspace\else, GordonJ03\fi}.
%
A trace property is a safety property if, within any (possibly infinite) trace
that violates the property, there exists a finite ``bad prefix'' that violates
it.
%
%
For instance here is a bad prefix for $S_2$ that includes a call to
\ls|E.write(<data,x>)| with no preceding call to \ls|C$_0$.valid(data)|:
\begin{lstlisting}
    [C$_0$.main();C$_2$.init();Ret;E.read;Ret(x);C$_1$.parse(x);
     Ret(y);C$_2$.process(y);Ret(data);E.write(<data,x>)]
\end{lstlisting}
The program from \autoref{fig:example-components-secure} cannot produce
traces with this bad prefix, but it could do so if we removed the validity
check in \ls|C$_0$.main()|; this variant would invalidate safety property
$S_2$.


%
\iflater
\bcp{On re-reading, I had exactly the same issue as Arthur (see
  comments).  Tried to do some rephrasing to clarify...}%
\fi
Compiler correctness is often phrased in terms of preserving
trace properties in general~\cite{Leroy09} (and thus safety properties
as a special case).
However, this \iffull notion of correctness \fi
is often predicated on the assumption
that the source program has no undefined behavior; if it does, all
security guarantees are lost, globally.
By contrast, we want our secure compilation criterion to still apply
even when some components are dynamically compromised by encountering
undefined behavior.
In particular, we want to ensure that dynamically
compromised components are not able to break the safety properties of
the system at the target level any more than equally privileged
components without undefined behavior already could in the source\iffull
language\fi.
\ifsooner
\bcp{Thinking like a new reader, I find the last sentence rather
  impenetrable.  ``We say compiler correctness is bad because it assumes
  that the source program has no undefined behavior; but then we reason
  about possible behaviors of compiled programs in terms of source programs
  with no undefined behaviors...???''  The point being made here seems very
  subtle, and I wonder whether we could make it more effectively if we
  postpone it till a bit later.  Or indeed whether it is too subtle to try
  to make at all, at least in this section.  (If we do move it, we'll need
  to repair the transition to the next paragraph a little...)}
\ch{Added the sentence in the middle for a clearer connection to CC}
\fi

\begin{figure}
\centering
\includegraphics[width=0.9\linewidth]{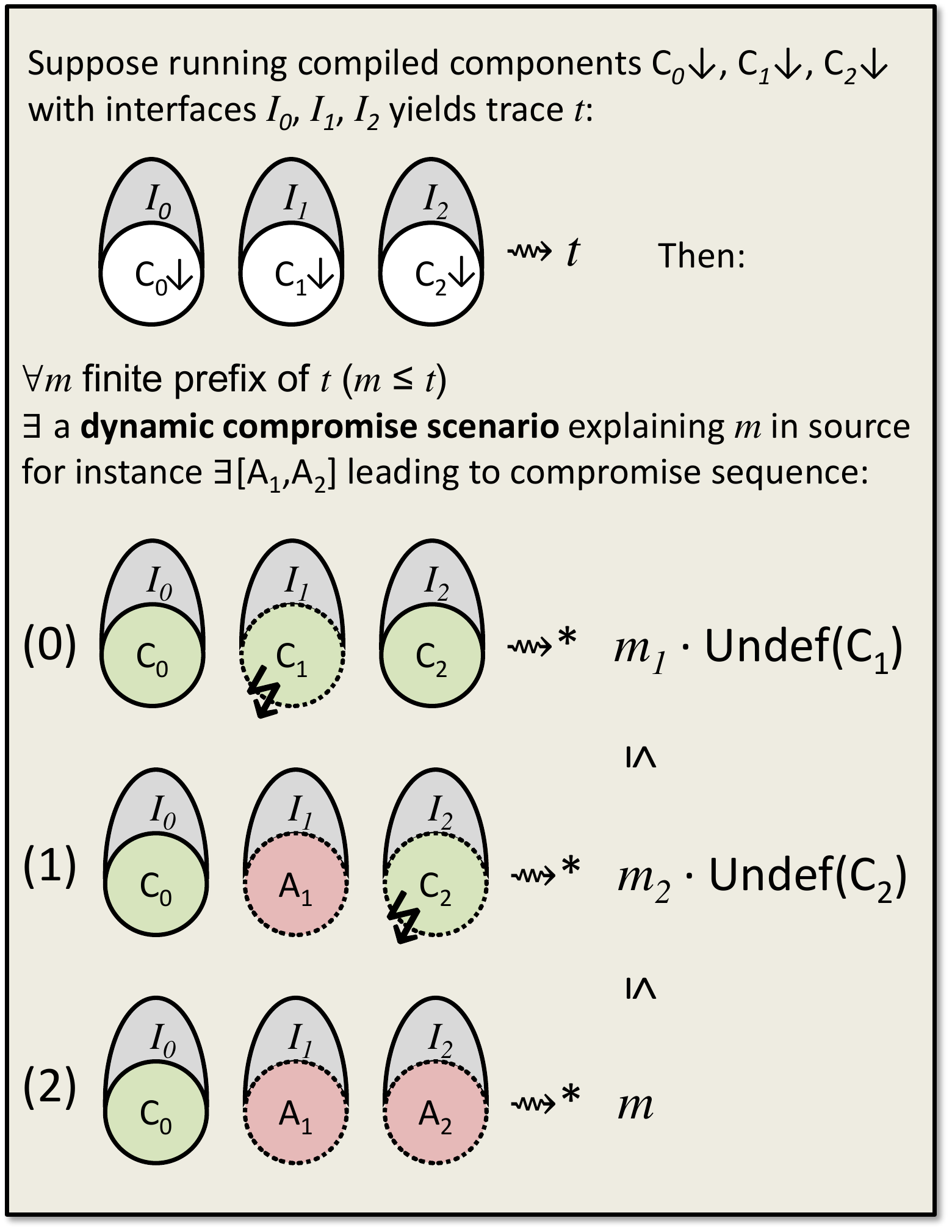}
\\[.4ex]
\begin{minipage}{0.9\linewidth}
The trace prefixes $m$, $m_1$, $m_2$ might, for instance, be:
\begin{lstlisting}
    $m$  = [C$_0$.main();C$_2$.init();Ret;E.read;Ret(x);C$_1$.parse(x);
          Ret(y);C$_2$.process(y);Ret(d);
          C$_0$.valid(d);Ret(true);E.write(<d,x>)]

    $m_1$ = [C$_0$.main();C$_2$.init();Ret;E.read;Ret(x);C$_1$.parse(x)]

    $m_2$ = [C$_0$.main();C$_2$.init();Ret;E.read;Ret(x);C$_1$.parse(x);
           Ret(y);C$_2$.process(y)]
\end{lstlisting}
\end{minipage}
\vspace{-0.8em}
\caption{The \rscc dynamic compromise game for our \iffull running \fi example.  We
  start with all components being uncompromised (in green) and incrementally
  replace any component that encounters undefined behavior with an arbitrary
  component (in red) that has the same interface and will do its part of the
  trace prefix $m$ without causing undefined behavior.\ifsooner\aaa{It took me a
    while to realize that the rotated $\leq$ was a rotated $\leq$, and not
    $I\Lambda$. Also, I couldn't understand what ``for instance,
    $\exists [A_1,A_2]$\ldots'' meant.}\bcp{+1 for both complaints, though
    I'm not sure what to do about the first.  For the second, people are
    likely to look at the diagram
    very early in their reading, so it's important that it stand on its own
    as much as possible.  It's worth burning a centimeter or so of vertical
    space to unpack the very telegraphic phrasing of the words in the
    diagram into something a bit easier to follow.}\fi
\ch{I have an improved variant of this in CCS talk!}
}
\label{fig:rcc-example}
\vspace{-1em}
\end{figure}

\ifsooner
\bcp{The next two paragraphs seem critical to the story, but I don't find
  them very clear.  It's a bit hard to articulate why, but maybe it has to
  do with the way the specific example is mixed with description of the
  general definition of \rscc.  I can have a go at rewriting, but that will
  probably be easier after some of the earlier comments are dealt with...}
\fi

We call our criterion {\em Robustly Safe Compartmentalizing Compilation}
(\rscc).  It is phrased in terms of a ``security game,''
\ifsooner\bcp{Is the ``game''
intuition important?  If so, then we need to be clearer about who
are the players, what are the moves, etc.  Otherwise trying to figure out
how to map these concepts onto the rest of the discussion is just
distracting for the reader.}\ch{what would be a better word for it?}\bcp{Why
do we need a word at all?  E.g., could we start out by saying something like
``Intuitively, it says that it should be possible to understand any trace of
the whole program---including a trace in which one or more
components become compromised at some point in the program's execution---in
terms of an alternative execution where the compromised components are
replaced by ones that are chosen arbitrarily by the attacker\ch{not really!}
but that do not
exhibit undefined behaviors''?}\ch{Sounds okay, but should distinguish
  target from source executions}\aaa{I agree with Benjamin}\fi
illustrated in \autoref{fig:rcc-example} for our running example.
\iflater
\aaa{This also sounds contradictory: we first say that we can explain \emph{any}
  prefix in terms of the source, but later add that this is not true in the
  presence of undefined behavior.}\ch{It only shows that one can't be too
  naive, but we do have a nice solution. Emphasized {\em directly} below,
  replace with naively if you so wish.}%
\fi
With an \rscc compilation chain,
given any execution of the compiled and linked components
\ls|C$_0\hspace{-0.35em}\downarrow$|,
\ls|C$_1\hspace{-0.35em}\downarrow$| and,
\ls|C$_2\hspace{-0.35em}\downarrow$| producing trace $t$ in the target language,
we can explain any (intuitively bad)
finite prefix $m$ of $t$ (written $m \leq t$) in terms of the source language.
As soon as any component of the program has an undefined behavior
though, the semantics of the source language can no longer {\em directly} help us.
Similar to CompCert~\cite{Leroy09}, we model undefined behavior in our source
language as a special event \ls|Undef(C$_i$)| that terminates the
trace.\iflater\aaa{Maybe ``inspired by'' would be more accurate, given that CompCert
  undefined behavior does not mention components?}\bcp{Seems OK to me
  as-is...}\fi{}
For instance, in step 0 of 
\autoref{fig:rcc-example}, component \ls|C$_1$| is the first to encounter
undefined behavior after producing a prefix $m_1$ of $m$.

Since undefined behavior can manifest as arbitrary target-level behavior,
the further actions of component \ls|C$_1$|
can no longer be explained in terms of its source code.  So how can we
explain the rest of $m$ in the source language?
Our solution in \rscc is to require that one can replace
\ls|C$_1$|, the component that encountered undefined behavior, with
some other source component \ls|A$_1$| that has the same interface and
can produce its part of the whole $m$ in the source language without itself
encountering undefined behavior.
In order to replace component \ls|C$_1$| with \ls|A$_1$| we
have to go back in time and re-execute the program from the beginning
obtaining a longer trace, in this case $m_2\cdot$\ls|Undef(C$_2$)|
(where we write ``$\cdot$'' for appending the
event \ls|Undef(C$_2$)| to $m_2$).
We iterate this process until all components that encountered undefined
behavior have been replaced with new source components that do not
encounter undefined behavior and produce the whole $m$.
In the example dynamic compromise scenario from \autoref{fig:rcc-example},
this means replacing
\ls|C$_1$| with \ls|A$_1$| and \ls|C$_2$| with \ls|A$_2$|, after which
the program can produce the whole prefix $m$ in the source\iffull language\fi.

Let's now use this \rscc security game to deduce that in our example from
\autoref{fig:example-components-secure}, even compromising both
\ls|C$_1$| and \ls|C$_2$| does not break property $S_2$ at the target level.
Assume, for the sake of a contradiction, that a trace of our compiled program breaks
property $S_2$.
Then there exists a finite prefix ``$m\mathop{\cdot}\text{\ls|E.write(<data,x>)|}$''
such that \ls|C$_0$.valid(data)| does not appear in $m$.
Using \rscc we obtain that there exists some dynamic compromise scenario
explaining $m$ in the source.
The simplest case is when no components are compromised.
The most interesting case is when this scenario involves the
compromise of both \ls|C$_1$| and \ls|C$_2$| as in \autoref{fig:rcc-example}.
%
In this case, replacing
\ls|C$_1$| and \ls|C$_2$| with arbitrary
\ls|A$_1$| and \ls|A$_2$| with the same interfaces allows us to
reproduce the whole bad prefix $m$ in the source (step 2
from \autoref{fig:rcc-example}).
We can now reason in the source, either informally or using a program
logic for robust safety \cite{SwaseyGD17}, that this cannot happen,
since the source code of \ls|C$_0$| does call \ls|C$_0$.valid(data)|
and only if it gets \ls|true| back does it call \ls|E.write(<data,x>)|.

While in this special case we have only used the last step in the
dynamic compromise sequence, where all compromised components have
already been replaced
(step 2 from \autoref{fig:rcc-example}), the previous steps are also
useful in general for reasoning about the code our original components
execute {\em before} they get \iffull dynamically \fi compromised.
For instance, this kind of reasoning is crucial for showing
property $W_2$ for the original example from \autoref{fig:example-components}.
Property $W_2$ gives up on the validity of the written data only if
\ls|C$_2$| receives a \ls|y| that exploits
\ls|C$_2$.handle(y)| (vulnerability $V_3$).
However, as discussed above, a compromised \ls|C$_1$|
could, in theory, try to compromise \ls|C$_2$| by
calling \ls|C$_2$.process| without proper initialization
(exploiting vulnerability $V_2$).
Showing that this cannot actually happen requires using step 0
of the game from \autoref{fig:rcc-example}, which gives us that the
original compiled program obtained by linking
\ls|C$_0\hspace{-0.35em}\downarrow$|,
\ls|C$_1\hspace{-0.35em}\downarrow$| and,
\ls|C$_2\hspace{-0.35em}\downarrow$| can produce the trace
$m_1\mathop{\cdot}\text{\ls|Undef(C$_1$)|}$, for some prefix $m_1$ of the bad
trace prefix in which \ls|C$_2$.process| is called without
calling \ls|C$_2$.init| first.
But it is easy to check that
the straight-line code of the \ls|C$_1$.main()| procedure can only
cause undefined behavior {\em after} it has called \ls|C$_2$.init|,
contradicting the existence of a bad trace
exploiting $V_2$.

\ch{Don't think we're making this explicit anywhere now, but the way
  our property is stated, it assumes determinacy. Otherwise, it
  wouldn't be possible to rewind execution along the same trace,
  since internal nondeterminism could mess things up.}



\ifsooner
\aaa{Thinking about it again, I find the discussion about source-level reasoning
  a bit misleading, because it can give the impression that what matters is the
  fact that the attackers are written in a particular language---e.g., ``given
  that the attacker code is inductively generated by this grammar, I conclude
  that...''  I think that what really matters is the fact the evolution of the
  state of the protected component is solely determined (1) by that component's
  (not the context's) code, and (2) the context's calls and returns into it.}
\bcp{+2}
\ch{I don't agree with this assessment. All our criteria are clearly
  stated in terms of syntax. You seem to be describing something else
  that might also be reasonable, but that has little to what we
  propose in this paper. Reminds me of discussions with Steven
  Schaefer, who thinks that full abstraction should be defined in
  terms of semantic not syntactic observers. That's a quite different
  security criterion though.
 \fi

\ifcheckpages\clearpage\fi
\section{Formally Defining \rscc}
\label{sec:formal}

\iflater
\ch{Here's another way we could explain the beginning of this
  section, in terms of compiler correctness:
(0) start from standard compiler correctness for safe languages
    and castrate it so that it only talks about safety properties;
(1) \rsc is an extension of castrated compiler correctness to
     partial programs that run in an adversarial low-level context;
(2) (whole program) castrated compiler correctness can also be
     extended to unsafe languages, as CompCert already does;
(3) Putting together (1) and (2) we get \rscdc, now that looks trivial;
... So this would take more space (explaining 2 things that are not
new to this work), make \rscdc look trivial, and make safety look stupid.
It might allow non-experts to understand this section more easily though.}
\ch{Did a simpler things and added a paragraph in 3.2 relating to a
    CompCert top-level theorem}
\fi


For pedagogical purposes, we define \rscc in stages, incrementally adapting the
existing notion of {\em Robustly Safe Compilation} (\rsc) introduced by
Abate~\ETAL\cite{AbateBGHPT19} (and reviewed in \autoref{sec:rsc}).  We first bring
\rsc to {\em unsafe languages with undefined behavior}
(\autoref{sec:rsc-dc}), and then further extend its protection to any set of
{\em mutually distrustful components} (\autoref{sec:rsc-dc-md}).  These ideas
lead to the more elaborate \rscc property (\autoref{sec:rcc-formal}), which
directly captures the informal dynamic compromise game from
\autoref{sec:rcc-informal}.  These definitions are generic, and will be
illustrated with a concrete instance in \autoref{sec:compiler}.
Finally, we describe an effective and general
proof technique for \rscc (\autoref{sec:proof-technique}).

\ifsooner
\apt{Maybe lift the description of the ``whole compilation chain''
  from second para. of \autoref{sec:rsc} to here and be more explicit
  that these are parameters of entire development.}
\fi



\subsection{\rsc\/: Robustly Safe Compilation}
\label{sec:rsc}

%
\ifsooner \bcp{The flow in this paragraph is a mess:}\fi
\rsc~\cite{AbateBGHPT19} is a recent criterion for secure compilation that
captures the preservation of all {\em robust} safety properties---\IE
safety properties that hold in the presence of arbitrary adversarial
contexts~\cite{KupfermanV99, GordonJ04, SwaseyGD17}.
%
%
A trace property (\IE{} a set of potentially infinite traces built over events
like I/O with the environment~\cite{Leroy09}) is a safety
property~\cite{LamportS84} if any trace violating it has a finite ``bad prefix''
that already violates it.
We focus on robust safety since it captures many important program properties
(\EG robust partial correctness\iffull~\cite{SwaseyGD17}\fi),
while allowing for a simple
secure-compilation proof technique (\autoref{sec:proof-technique}).

\rsc is a property of a whole compilation chain: the source
language and its trace-based big-step operational semantics (we write
$P {\rightsquigarrow} t$ to mean that the complete program $P$ can
produce trace $t$), plus its compiler ($\cmp{P}$), source and target linking
(where $C_S[P]$ denotes linking a partial program $P$ with context
$C_S$ to obtain a whole source program, and $C_T[P_T]$ does the same
in the target), and target-level semantics ($P_T {\rightsquigarrow} t$)
\iflater
\aaa{I wonder if introducing this notation here makes things more confusing
  instead of less. I think it would be clearer if we just defined everything
  after the formula in \autoref{defn:rsc} with a ``where'' list}\ch{I like
  it better before and can't see any confusion.}%
\fi
including for instance the target machine, loader, and deployed protection
mechanisms.
\begin{defn}\label{defn:rsc}
A compilation chain provides \ifsooner\bcp{we could save a little space by putting
  less padding above and below displayed equations --- this one actually
  looks funny because it has so much!}\ch{+1}\fi
\iffull{\em Robustly Safe Compilation (\fi\rsc\iffull)}\fi iff
\[
\forall P~C_T~t.~
  C_T[\cmp{P}] {\rightsquigarrow} t \Rightarrow
  \forall m {\leq} t.~ \exists C_S~t'.~  C_S[P] {\rightsquigarrow} t' \wedge m {\leq} t'.
\]
\end{defn}

That is, \rsc holds for a compilation chain if,
for any partial source program $P$ and any target
context $C_T$, where $C_T$ linked with the compilation of $P$
can produce a trace $t$ in the target \iffull language\fi
($C_T[\cmp{P}] {\rightsquigarrow} t$),
and for any finite prefix $m$ of trace $t$ (written $m \leq t$),
we can construct a
source-level context $C_S$ that can produce prefix $m$ in the source
language when linked with $P$ (\IE $ C_S[P] {\rightsquigarrow} t'$ for some
$t'$ so that $m \leq t'$).\iflater\bcp{Oof!}\fi{}
Intuitively, if we think of the contexts as
adversarial and $m$ as a bad behavior, \rsc says
that any finite attack $m$ that a target context $C_T$ can mount against
$\cmp{P}$ can already be mounted against $P$ by some source context $C_S$.
So proving \rsc requires that we be able to {\em back-translate} each
finite prefix $m$ of $C_T[\cmp{P}]$ into a source context $C_S$ that
performs $m$ together with the original program $P$.
Conversely, any safety property that holds of $P$ when linked with an
arbitrary source context will still hold for $\cmp{P}$ when linked
with an arbitrary target context~\cite{AbateBGHPT19}.

As in CompCert, we assume \iffull for simplicity \fi that the traces are
exactly the same in the source and target languages. We anticipate no
trouble relaxing this to an arbitrary relation between source and target traces.




\subsection{\rscdc\/: Dynamic Compromise}
\label{sec:rsc-dc}

\ifsooner\ch{Might want to make it
  clear upfront, maybe even by an example in the previous subsection why
  \rsc doesn't work for an unsafe language.\apt{+1} Now it should only become
  clearer below, but it also gets technical.}

\aaa{I tried to explain the limitation of \rsc, but I got confused: shouldn't
  \rsc hold for a language that modeled the occurrence of undefined behavior as
  non-determinism in the source?}\ch{see slack}

\ch{This seems to explain why compiler correctness for safe
  languages doesn't apply to unsafe ones. At this point only
  RSC is in scope.}%
When applied to a complete program $P$, \rsc roughly implies that any trace
produced by the compiled program $\cmp{P}$ could have been produced at the
source level.  This requirement does not hold for compilation chains for
languages with undefined behavior.
For example, if a C program has undefined behavior, its compilation
might perform a system call that is never mentioned in the program text.
To overcome this limitation, we adapt \rsc to accommodate {\em unsafe} source
languages, in which the protected partial program itself can become compromised.
\fi{}

The \rsc criterion above
is about protecting a partial program written in a {\em safe} source
language against adversarial target-level contexts.
We now adapt the idea behind \rsc to an {\em unsafe} source language
with undefined behavior, in which the protected partial program itself
can become compromised.
As explained in \autoref{sec:rcc-informal}, we model undefined behavior as a
special \ls|Undef| event terminating the trace: whatever happens afterwards at
the target level can no longer be explained in terms of the code of the source
program.
We further assume that each undefined behavior in the
\iffull semantics of the \fi source
language can be attributed to the part of the program that causes it
by labeling the \ls|Undef| event with ``blame the program'' ($P$) or
``blame the context'' ($C$)
(while in \autoref{sec:rsc-dc-md}
we will blame the precise component encountering undefined behavior).
%
%
\begin{defn}\label{defn:rsc-dc}
  A compilation chain provides {\em Robustly Safe Compilation
    with Dynamic Compromise (\rscdc)} iff
%
\[
\forall P~C_T~t.~
  C_T[\cmp{P}] {\rightsquigarrow} t \Rightarrow
  \forall m{\leq}t.~
  \exists C_S~t'.~ C_S[P] {\rightsquigarrow} t'
  \wedge (m {\leq} t' {\color{red} \vee t' {\prec_P} m})
  .
\]
\end{defn}
Roughly, this definition relaxes \rsc by forgoing protection for the partial
program $P$ after it encounters undefined behavior.  More precisely, instead of
always requiring that the trace $t'$ produced by $C_S[P]$ contain the entire
prefix $m$ (\IE $m {\leq} t'$), we also allow $t'$ to be itself a prefix of $m$
followed by an undefined behavior in $P$, which we write as $t' {\prec_P} m$
(\IE
$t' {\prec_P} m \triangleq \exists m'{\leq}m.~ t'{=}(m'\cdot\texttt{Undef}(P))$).
In particular, the context $C_S$ is guaranteed to be free of undefined
behavior before the whole prefix $m$ is produced
or $P$ encounters undefined behavior.
However, nothing prevents $C_S$ from passing values to $P$ that
try to trick $P$ into causing undefined behavior.

To illustrate, consider the partial program $P$ defined below.
\iffull
It exports a
single function \ls|foo| that processes its argument and writes out the result.
\fi
\iffull
\bcp{Can we make the font in these examples any bigger?  Right now it's a bit
  hard on the eyes and looks funny on the page.}
\fi

\noindent
\begin{minipage}{.65\columnwidth}
\begin{lstlisting}
program $P$ {
  import E.write; export foo;
  foo(x) {
    y := P.process(x);
    E.write(y);
  }
  // can encounter Undef for some x
  process(x) { ... }
}
\end{lstlisting}
\end{minipage}
\begin{minipage}{.35\columnwidth}
\begin{lstlisting}
context $C_S$ {
  import E.read, $P$.foo;
  main() {
    x := E.read();
    $P$.foo(x);
  }
}
\end{lstlisting}
\end{minipage}
Suppose we compile $P$ with a compilation chain that satisfies \rscdc, link the
result with a target context $C_T$ obtaining $C_T[\cmp{P}]$, execute this and
observe the following finite trace prefix:
\begin{lstlisting}
$m$ = [E.read(); Ret("feedbeef"); P.foo("feedbeef"); E.write("bad")]
\end{lstlisting}
According to \rscdc there exists a source-level context $C_S$ (for
instance the one above) that explains the prefix $m$ in terms of the
source language in one of two ways:
either $C_S[P]$ can do the entire $m$ in the source,
or $C_S[P]$ encounters an undefined behavior in $P$ after a prefix of $m$,
for instance the following one:
\begin{lstlisting}
$t'$ = [E.read(); Ret("feedbeef"); P.foo("feedbeef"); Undef(P)]
\end{lstlisting}

\ifsooner
\ch{This example is simple, but I'm starting to wonder if it's too
  simple. In dynamic-compromise.org I have another example that
  tries to illustrate reasoning with RSC_DC in the style from
  section 2. That would take a lot more space though, so unsure.}

\ch{William (\#W4): Shortly before section 3.3, I began to wonder:
  "okay, so what? if I'm preserving source-level reasoning, *kind of*,
  up to undefined behavior... how does this help me as a programmer?"}
\fi


As in CompCert~\cite{Regehr10, Leroy09}, we treat undefined
behaviors as {\em observable} events at the end of the execution trace,
allowing compiler optimizations that move an undefined
behavior to an earlier point in the execution, but not past any
other observable event.
While some other C compilers would need to be adapted to respect this
discipline~\cite{Regehr10}, limiting the temporal scope of undefined
behavior is a necessary prerequisite for achieving security against
dynamic compromise.
Moreover, if trace events are coarse enough (\EG system calls and
cross-component calls) we expect this restriction to have a negligible
performance impact in practice.

\iflater
\ch{David Chisnall pointed out in Dagstuhl that this constraint would
  be hard to enforce in a concurrent setting, where, according to
  David, a racy program does have undefined behavior globally. He even
  seemed to imply that even the order of events in a single thread
  might not be respected for the most crazy concurrency models with
  ``out of thin air reads''. Will need to keep all this in mind if we
  ever want to extend what we do with concurrency.}
\ch{Later discussed with Jeehoon Kang, who told me that the community
  is moving towards making data races defined behavior, and their weak
  memory model/semantics definitely goes in that direction. So the
  perspectives are not as bad as David seemed to imply. This might mean
  fixing the C standard though.}
\fi

%

One of the top-level CompCert theorems\iffull\footnote{Theorem
{\tt transf\_c\_program\_preservation} in file
{\tt \href{https://github.com/AbsInt/CompCert/blob/master/driver/Complements.v\#L41}{Complements.v}}, which is a corollary of the main semantic preservation theorem
of CompCert.
}\fi{}
does, in fact, already capture dynamic compromise in a similar way to \rscdc.
Using our notations this CompCert theorem looks as follows:
\[
\begin{array}{l}
\forall P~t.~
  (\cmp{P}) {\rightsquigarrow} t \Rightarrow
  \exists t'.~ P {\rightsquigarrow} t' \wedge (t' {=} t \vee t' {\prec} t)
\end{array}
\]
This says that if a compiled whole program $\cmp{P}$ can produce a
trace $t$ with respect to the target semantics, then in the source $P$
can produce either the same trace or a prefix of $t$ followed by
undefined behavior.
In particular this theorem does provide guarantees to undefined
programs up to the point at which they encounter undefined behavior.
%
The key difference compared to our secure compilation chains
is that CompCert does not restrict undefined behavior
{\em spatially}: in CompCert undefined behavior
breaks all security guarantees of the {\em whole} program, while in
our work we restrict undefined behavior to the component that causes it.
This should become clearer in the next section, where we explicitly
introduce components, but even in \rscdc we can already imagine
$\cmp{P}$ as a set of uncompromised components for trace prefix $m$,
and $C_T$ as a set of already compromised ones.

A smaller difference with respect to the CompCert theorem is that
(like \rsc) \rscdc only looks at finite prefixes in order to simplify
the difficult proof step of context back-translation,
which is not a concern that appears in CompCert and
the usual verified compilers\ifsooner\apt{such as? cite or omit}\fi.
\autoref{sec:zp-details} precisely characterizes the subclass of
safety properties that is preserved by \rscdc even in adversarial contexts.

\subsection{\rscdcmd\/: Mutually Distrustful Components}
\label{sec:rsc-dc-md}

\rscdc gives a model of {dynamic compromise} for
secure compilation, but is still phrased in terms of protecting a
trusted partial program from an untrusted context.
We now adapt this model to protect any set of {\em mutually distrustful
  components} with clearly specified privileges from an untrusted context.
Following Juglaret~\ETAL's work in the full abstraction
setting~\cite{JuglaretHAEP16}, we start by taking
both partial programs and contexts to be sets of components;
linking a program with a context is then just set union.
We compile sets of components by separately compiling each component.
Each component is assigned a well-defined interface that precisely
captures its {\em privilege}; components can only interact
as specified by their interfaces.
Most importantly, context back-translation respects these interfaces:
each component of the target context is mapped back to a source
component with exactly the same interface.
As Juglaret~\ETAL{} argue, \iffull the whole idea of \fi least-privilege design
crucially relies on the fact that, when a component is compromised,
it does not gain any more privileges.


\begin{defn}\label{defn:rsc-dc-md}
A compilation chain provides {\em Robustly Safe Compilation with
  Dynamic Compromise and Mutual Distrust (\rscdcmd)} if
there exists a back-translation function $\uparrow$ taking a finite
trace prefix $m$ and a component interface $I_i$ to a source component
with the same interface, such that, for any compatible interfaces $I_P$
and $I_C$,
\[
\begin{array}{l}
\forall P {:} I_P.~ \forall C_T{:}I_C.~ \forall t.~
  (C_T \cup \cmp{P}) {\rightsquigarrow} t \Rightarrow
  \forall m{\leq}t.~\\\quad
  \exists t'.~ (\{\back{(m,I_i)} \,|\, I_i \in I_C \} \cup P) {\rightsquigarrow} t'
  \wedge (m {\leq} t' \vee t' {\prec_{I_P}} m)
  .
\end{array}
\]
\end{defn}
This definition closely follows \rscdc, but it restricts
programs and contexts to compatible interfaces $I_P$ and $I_C$.
We write $P:I$ to mean ``partial program $P$ satisfies interface $I$.''
%
The source-level context is obtained by applying the back-translation
function $\uparrow$ pointwise to all the interfaces in $I_C$.
As before, if the prefix $m$ is cropped prematurely because of an
undefined behavior, then this undefined behavior must be in one of the program
components, not in the back-translated context components ($t' {\prec_{I_P}}
m$).

\iffull
\aaa{Could we write $t'  \prec_P m$ instead, for consistency with the previous
  section?}\ch{If we want consistency we should do the opposite, make 2 more
  precise not 3 less}

\aaa{Question: is \rscdcmd morally the same as \rscdc plus the fact that
  target-level traces respect interfaces?}\ch{This misses the most important
  part: it starts with ``Most importantly'' above}
\fi



\iflater
\ch{Warning: $C$ stands for both context and component, and both are
  standard, what can we do? Now $c$ is free, use it for components maybe?
  That would mean making all of them lower case though: $a$ for attacker?}
\ch{Warning: $c$ is also used to mean concrete in $\ii{cs}$ (concrete state)}
\fi


\iflater
\ch{The following is trying to extend robust preservation from \rscdc
  to \rscdcmd.  It should be a very unsurprising thing though.}
\ca{Assume we can replace the notion of robust preservation with
  the following $TP$ for $\pi$:
  \[
  (\exists C_T t. ~ C_T \cup P \downarrow \rightsquigarrow t \wedge t \not \in \pi) \Rightarrow
  (\exists t'. ~ (C_S \uparrow \cup P) \rightsquigarrow t' \wedge t \not \in \pi)
  \]
  where $\uparrow$ is a back translating function that preserves interfaces.
  Then with the same argument proved for $RSC^{DC}$ we can prove
  $\rscdcmd \Leftrightarrow (\forall P \pi \in Z_P. ~ TP ~P)$}
\fi

\subsection{Formalizing \rscc}
\label{sec:rcc-formal}

Using these ideas, we now define \rscc by following the
dynamic compromise game illustrated in \autoref{fig:rcc-example}.
We use the notation $P\doesprefix m$ when there exists a trace $t$
that extends $m$ (\IE $m \leq t$) such that $P {\rightsquigarrow} t$.
We start with all components being uncompromised and incrementally
replace each component that encounters undefined behavior in the source
with an arbitrary component with the same interface
that may now attack the remaining components.
\iffull Formally, this is captured by the following property: \fi
\begin{defn}\label{defn:rcc}
A compilation chain provides {\em Robustly Safe
  Compartmentalizing Compilation (\rscc)} iff
\iffull for any \else $\forall$\fi compatible interfaces $I_1, ..., I_n$,
%
%
\[
\begin{array}{l}
\forall C_1 {:} I_1,..., C_n {:} I_n.~\forall m.~
\{\cmp{C_1},...,\cmp{C_n}\} \doesprefix m \Rightarrow\\
  \exists A_{i_1} {:} I_{i_1},..., A_{i_k}{:}I_{i_k}.\\
~~~(1)~~\forall j \in 1 ... k.~ \exists m_j.~ (m_j \prec_{I_{i_j}} m) ~\wedge\iffull\\
\qquad (j{>}1 \Rightarrow\else(\fi m_{j-1} \prec_{I_{i_{j-1}}} m_{j}) ~\wedge\\
\quad\;\;\;\;  (\{C_1,...,C_n\} \backslash \{C_{i_1},...,C_{i_j{-}1}\} {\cup} \{A_{i_1},...,A_{i_j{-}1}\}) \doesprefix m_j\\[0.5em]
\wedge~(2)~~(\{C_1,...,C_n\} \backslash \{C_{i_1},...,C_{i_k}\} {\cup} \{A_{i_1},...,A_{i_k}\}) \doesprefix m
.
\end{array}
\]
\end{defn}
%
This says that $C_{i_1},...,C_{i_k}$ constitutes
a compromise sequence corresponding to finite prefix $m$
produced by a compiled set of components $\{\cmp{C_1},...,\cmp{C_n}\}$.
In this compromise
sequence each component $C_{i_j}$ is taken over by the already
compromised components at that point in time
$\{A_{i_1},...,A_{i_j{-}1}\}$ (part $1$).
%
%
Moreover, after replacing all the compromised components
$\{C_{i_1},...,C_{i_k}\}$ with their corresponding source components
$\{A_{i_1},...,A_{i_k}\}$ the entire $m$ can be reproduced in
the source language (part~$2$).

This formal definition allows us to play an iterative game in which
components that encounter undefined behavior successively become compromised and
attack the other components.
This is the first security definition in this space to support both
dynamic compromise and mutual distrust, whose interaction is subtle
and has eluded previous attempts at characterizing the security
guarantees of compartmentalizing compilation as extensions of fully
abstract compilation~\cite{JuglaretHAEP16}
(further discussed in \autoref{sec:related}).

\iflater
\ch{One thing we were wondering is whether the non-rewinding variant
  of RSCC can in fact still be proved from the rewinding one, maybe
  under some extra assumptions. Q: Would it hold for our concrete instance?}
\fi

\subsection{A Generic Proof Technique for \rscc}
\label{sec:proof-technique}

We now describe an effective and general proof technique for \rscc.
First, we observe that the slightly simpler
\rscdcmd implies \rscc.
Then we provide a generic proof in Coq that a compilation chain
obeys \rscdcmd if it satisfies certain well-specified assumptions on
the source and target languages and the \iffull compartmentalizing \fi
compilation chain.
%
%

Our proof technique yields simpler and more scalable proofs than
previous work in this space~\cite{JuglaretHAEP16}.
In particular, it allows us to directly reuse
a compiler correctness result {\em \`a la} CompCert,
which supports separate compilation but only guarantees correctness
for whole programs~\cite{KangKHDV15}; which avoids proving any
other simulations between the source and target languages.
Achieving this introduces some slight complications in the proof
structure, but it nicely separates the
correctness and security proofs and allows us to more easily tap
into the CompCert infrastructure.
Moreover, this novel proof structure completely removes the need for any kind of
``partial semantics''~\cite{JuglaretHAEP16, AbateBGHPT19} (a.k.a. ``trace
semantics''~\cite{JeffreyR05, PatrignaniC15}), allowing the proof to be done
entirely with respect to the semantics of {\em whole} programs in the source and
target languages.
Finally, since only the last step of our proof technique is
specific to unsafe languages, our technique could be further simplified
to provide scalable proofs of vanilla \rsc
for safe source languages~\cite{AbateBGHPT19, PatrignaniG18}.

\myindentedpar{\rscdcmd implies \rscc}
\label{sec:rscdc-implies-rcc}
The first step in our proof technique reduces \rscc to \rscdcmd,
using a theorem showing that
\rscc can be obtained by iteratively applying \rscdcmd.
This result crucially relies on back-translation in \rscdcmd being
performed pointwise and respecting interfaces,
as explained in \autoref{sec:rsc-dc-md}.

\begin{thm}\label{thm:rscdcmd-rcc}
\rscdcmd implies \rscc.
\end{thm}
We proved this by defining a non-constructive function that produces
the compromise sequence $A_{i_1}, ..., A_{i_1}$ by case analysis on
the disjunction in the conclusion of \rscdcmd (using excluded middle
in classical logic). If $m \leq t'$ we are
done and we return the sequence we accumulated so far, while if
$t' {\prec_P} m$ we obtain a new compromised component $c_i:I_i$ that we
back-translate using $(m,I_i)\uparrow$ and add to the sequence before
iterating this process.


\ifsooner
\bcp{Is ``classic RS'' a degenerate case of RSCC?}
\ch{{\em RCC/RSCC} is an iterative variant of \rscdcmd.  I'm not sure
  whether {\em RCC/RSCC} implies \rscdcmd.  I think it's possible, but
  would need to actually prove it before claiming anything.  At this
  point all we have + needed is the opposite direction
  (\autoref{thm:rscdcmd-rcc})}\bcp{Might it help the reader's intuition to
  at least make a conjecture about this?}
\ch{We would need to first make the exists in RSCC
  be more constructive, since classical logic would probably produce
  non-computable functions with too many arguments.
\[\footnotesize
\begin{array}{l}
\forall C_1 {:} I_1,..., C_n {:} I_n.~\forall m.~
\{\cmp{C_1},...,\cmp{C_n}\} \doesprefix m \Rightarrow
  \exists {i_1},..., {i_k}.\\
(1)~\forall j \in 1 ... k.~ \exists m_j.~ (m_j \prec_{I_{i_j}} m) ~\wedge\iffull\\
\qquad (j{>}1 \Rightarrow\else(\fi m_{j-1} \prec_{I_{i_{j-1}}} m_{j}) ~\wedge\\
\quad\;  (\{C_1,...,C_n\} \backslash \{C_{i_1},...,C_{i_j{-}1}\}
  {\cup} \{\back{(m,I_{i_1})},...,\back{(m,I_{i_j{-}1})}\}) \doesprefix m_j\\[0.5em]
(2)~(\{C_1,...,C_n\} \backslash \{C_{i_1},...,C_{i_k}\}
  {\cup} \{\back{(m,I_{i_1})},...,\back{(m,I_{i_k})}\}) \doesprefix m
\end{array}
\]
  Then, since RSCC doesn't have a notion of program vs context we have
  to do case analysis on the compromise sequence. If it contains no
  component from $P$ then we need to produce the whole trace (RSCC part
  2), while otherwise we can stop at a prefix (RSCC part 1, the step in
  which the first program component has undefined behavior). For the
  back-translation we would use the $\uparrow$ function on the
  interfaces of context components in the compromise sequence and
  identity for the uncompromised context components. So it can probably
  work, although it requires changes and
  I'm not super confident at this point.}
\fi

\ifsooner
\bcp{By the way, I'm realizing that I've been using iffull, ifsooner, and
  iflater more or less interchangeably over the past couple of days.  At
  some point we should turn on ALL these flags and tidy / rationalize.}
\fi

\myindentedpar{Generic \rscdcmd proof outline}
Our high-level \rscdcmd proof is generic and works for any
compilation chain that satisfies certain well-specified
assumptions, which we introduce informally for now, leaving
details to the end of this sub-section.
The \rscdcmd \iffull(and thus \rscc)\fi proof for the compiler chain in
\autoref{sec:compiler} proves all these assumptions.\ifsooner\ch{Trying
  to imply no completeness}\fi{}
%

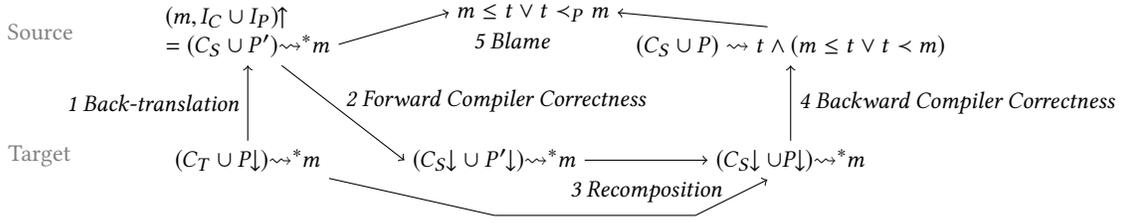
\begin{figure*}
\centering
\begin{tikzpicture}[auto]
  \node(cPi) {$(C_T \cup \cmp{P}) \doesprefix m$ };
  \node[right = of cPi] (CP1t) { $(\cmp{C_S} \cup~ \cmp{P'}) \doesprefix m$ };
  \node[align = left, above = of cPi] (CP1s) { $\back{(m, I_C \cup I_P)}$ \\ $= (C_S \cup P') \doesprefix m$ };
  \node[right = of CP1t, xshift=2em] (CPt) { $(\cmp{C_S} \cup \cmp{P}) \doesprefix m$ };
  \node[above = of cPi, xshift=12em, yshift=1.5em] (prec) { $m \leq t \lor t \prec_P m$ };
  \node[above = of CPt] (CPs) { $(C_S \cup P) \rightsquigarrow t \wedge (m \leq t \lor t \prec m)$};

  \draw[->] (cPi.90) to node {\hyperref[asm:bt]{\em 1 Back-translation}} (CP1s.-90);

  \draw[->] (CP1s.-45) to node [align=left,font=\itshape,xshift=-0.25em,yshift=-0.3em]{\hyperref[asm:fcc]{2 Forward Compiler Correctness}} (CP1t.180);

  \draw[->] (CP1s.-8) to node [right,yshift=-0.5em, xshift=3em]{\em \hyperref[asm:blame]{5 Blame}} (prec.180);
  \draw[->] (CPs.140) to node {} (prec.0);

  \draw[->] (CP1t.0) to node [below,xshift=0em,yshift=-0.5em]{\em \hyperref[asm:composition]{3 Recomposition}} (CPt.180);
  \coordinate [below=1.5em of CP1t,xshift=0em] (compoint1);
  \coordinate [below=1.5em of CPt,xshift=-4em] (compoint2);
  \draw[-] (cPi) to (compoint1);
  \draw[-] (compoint1) to (compoint2);
  \draw[->] (compoint2) to (CPt.220);

  \draw[->] (CPt.90) to node [right,align=left,font=\itshape]{\hyperref[asm:bcc]{4 Backward Compiler Correctness}} (CPs.-90);

  \node[left = of CP1s, xshift=0em] () {\color{gray} Source};
  \node[left = of cPi, xshift=-0.5em,yshift=0.2em] () {\color{gray} Target};
\end{tikzpicture}
\vspace{-0.8em}
\caption{Outline of our generic proof technique for \rscdcmd}
\label{fig:rsc-dc-md-proof}
\end{figure*}

The proof outline is shown in \autoref{fig:rsc-dc-md-proof}.
We start (in the bottom left) with a complete target-level program $C_T \cup
\comp{P}$
producing a trace with a finite prefix $m$ that we assume contains
no undefined behavior (since we expect that the final target
of our compilation will be a machine for which all behavior is defined).\ch{The
  POPL'20 reviewer shattered this dream of ours, so we might need to adapt one day.}
%
The prefix $m$ is first back-translated to synthesize a complete source program
$C_S \cup P'$ producing $m$ (the existence and correctness of this
back-translation are \autoref{asm:bt}).
For example, for the compiler in \autoref{sec:compiler}, each component $C_i$
produced by back-translation uses a private counter
to track how many events it has produced
during execution.  Whenever $C_i$ receives control, following an external call or
return, it checks this counter to decide what event to emit next, based on the
order of its events on $m$
(see \autoref{sec:rscc-theorem} for details).

The generated source program $C_S \cup P'$ is then separately compiled to a
target program $\cmp{C_S} \cup~ \cmp{P'}$ that, by compiler correctness,
produces again the same prefix $m$ (\autoref{asm:fcc}).
%
Now from $(C_T \cup \cmp{P}) \doesprefix m$ and $(\cmp{C_S} \cup~ \cmp{P'})
\doesprefix m$ we would like to obtain $(\cmp{C_S} \cup~ \cmp{P}) \doesprefix m$
by ``recomposing'' (\autoref{asm:recomposition}) the components
from the two executions and
forming a new execution for $\cmp{C_S} \cup~ \cmp{P}$.
We use a three-way simulation to relate the executions of the two complete
programs, to a third execution that combines the program part of the first
execution, and the context part of the second execution.

%
%
%
%

Once we know that $(\cmp{C_S} \cup~ \cmp{P}) \doesprefix m$, we use
compiler correctness again---now in the backwards direction
(\autoref{asm:bcc})---%
to obtain an execution of the source program $C_S \cup P$ producing trace $t$.
Because our source language is unsafe, however, \iffull trace \fi $t$ need not be
an extension of $m$: it can end earlier with an undefined behavior
(\iffull as also discussed in \fi\autoref{sec:rsc-dc}%
).
So the final step in our proof shows that if the source execution ends
earlier with an undefined behavior ($t' {\prec} m$),
then this undefined behavior can only be caused by $P$
(\IE $t' {\prec}_P m$), not by $C_S$, which was correctly generated
by our back-translation (\autoref{asm:blame}).
%
%


\myindentedpar{Assumptions of the \rscdcmd proof}
The generic \rscdcmd proof outlined above relies on
assumptions about the compartmentalizing compilation chain.
In the reminder of this subsection we give details about these
assumptions, while still trying to stay at a high level by omitting some of
the low-level details in our Coq formalization.
%

The first assumption we used in the proof above is that every
trace prefix that a target program can produce can also be produced by
a source program with the same interface.
A bit more formally, we assume the existence of
a \emph{back-translation function}~ $\back{}$ that given
a finite prefix $m$ that can be produced by a whole target program $P_T$,
returns a whole source program with the same interface $I_P$ as $P_T$
and which can produce the same prefix $m$ (\IE~$\back{(m,I_P)} \doesprefix m$).
\begin{assumption}[Back-translation]\label{asm:bt}
\[
  \exists \uparrow.~ \forall P{:}I_P.~\forall m~\ii{defined}.~
  P \doesprefix m \Rightarrow
  \back{(m,I_P)} \mathop{:} I_P \mathrel{\wedge} \back{(m,I_P)} \doesprefix m
\]
\end{assumption}

Back-translating only finite prefixes enables our proof technique, but at the same
time limits it to only safety properties.  While the other assumptions from this
section can probably also be proved for infinite traces, there is no general way
to define a finite program that produces an arbitrary infinite trace.  We leave
devising scalable back-translation proof techniques that go beyond safety
properties to future work\iffull (\autoref{sec:conclusion})\fi.

It is not always possible to take an {\em arbitrary}
finite sequence of events and obtain a source program that realizes it.
For example, in a language with a call stack and events
$\{\mathtt{call},\mathtt{return}\}$, there is no program that produces
the single event trace $\mathtt{return}$, since every return must be
preceded by a call.
Thus we only assume we can back-translate prefixes that are
produced by the target semantics.

As further discussed in \autoref{sec:related}, similar back-translation
techniques that start from finite execution prefixes
have been used to prove fully abstract compilation~\cite{JeffreyR05, PatrignaniC15}
and very recently \rsc~\cite{PatrignaniG18} and
even stronger variants~\cite{AbateBGHPT19}.
Our back-translation, on the other hand, produces not just a source context, but
a whole program.  In the top-left corner of \autoref{fig:rsc-dc-md-proof}, we
assume that this resulting program, $\back{(m, I_C \cup I_P)}$, can be
partitioned into a context $C_S$ that satisfies the interface $I_C$, and a
program $P'$ that satisfies $I_P$.\ch{To really obtain \rscdcmd we in
  fact need to further break up $C_S$ all the way to its individual
  components. This is only possible when back-translation works
  pointwise wrt components, which is an assumption we should make more
  explicit here, and should also properly expose in the Coq proofs.}

Our second assumption is a form of forward compiler correctness for
unsafe languages and a direct consequence of a forward simulation
proof in the style of CompCert~\cite{Leroy09}.
We assume separate compilation, in the style of a recent extension
proposed by Kang~\ETAL\cite{KangKHDV15} and implemented in CompCert since
version 2.7.
Our assumption says that if a whole program composed of \iffull parts \fi $P$ and
$C$ (written $C \cup P$) produces the finite trace prefix $m$ that does not
end with undefined behavior ($m~\ii{defined}$) then $P$ and $C$ when
separately compiled and linked together
($\cmp{C} \mathop{\cup} \cmp{P}$) can also produce $m$.\ch{This step crucially
  depends on m defined, but we might need to move away from that.}
\begin{assumption}[Forward Compiler Correctness with Separate Compilation and Undefined Behavior]\label{asm:fcc}
\[
  \forall C~P.~ \forall m~\ii{defined}.
  ~(C \cup P) \doesprefix m \Rightarrow (\cmp{C} \mathop{\cup} \cmp{P}) \doesprefix m
\]
\end{assumption}
%
%

The next assumption we make is {\em recomposition}, stating that if two
programs, each composed of a partial program side $\cmp{P}$ or $\cmp{P'}$, and a context
side $C_T$ or $\cmp{C_S}$, sharing the same interfaces, produce the same finite trace
prefix $m$, then the complete program $\cmp{C_S} \cup \cmp{P}$ obtained by linking one
side of each produce the same $m$.

\begin{assumption}[Recomposition]\label{asm:recomposition}
  \[
  \begin{array}{l}
    \forall P{:}I_P.~\forall P'{:}I_P.~ \forall C_T{:}I_C.~\forall C_S{:}I_C. ~
    \forall m. \\
    \quad (C_T \cup \cmp{P}){\rightsquigarrow}^* m \wedge
      (\cmp{C_S} \cup \cmp{P'}){\rightsquigarrow}^* m
    \Rightarrow
    (\cmp{C_S} \cup \cmp{P}){\rightsquigarrow}^* m
  \end{array}
  \]
\end{assumption}

In a previous version of this work~\cite{AbateABEFHLPST18}, we used to split
this theorem into two separate theorems, using a notion of partial semantics
\cite{JuglaretHAEP16, AbateBGHPT19} (a.k.a. ``trace
semantics''~\cite{JeffreyR05, PatrignaniC15}).
\jt{Could use $\cmp{P}$, etc. again below, but the compiler was not mentionned
  in the previosu version of the paper}
First, {\em decomposition} stated that if a program obtained by linking two
partial programs $P_T$ and $C_T$ produces a finite trace prefix $m$ that does
not end in an undefined behavior in the complete semantics, then each of the two
partial programs (below we take $P_T$, but the $C_T$ case is symmetric) can
produce $m$ in the partial semantics.
Then, the converse of decomposition, \emph{composition}, stated that if
two partial programs with matching interfaces produce the same prefix $m$ with
respect to the partial semantics, then they can be linked to produce the same
$m$ in the complete semantics.
When taken together, composition and decomposition captured the fact that
the partial semantics of the target language were adequate with respect to their
complete counterpart.
However, we noticed that proving those two assumptions independently
was particularly complex. Indeed, while decomposition is a straightforward
lemma provided the partial semantics are correctly setup, this is not the
case for composition.
Composing the two runs in the face of missing information was fraught
with delicate and tedious reasoning, and did not fit the standard simulation
mold we hoped it would.
This is why we instead propose to obtain the new run in the complete semantics
from the two runs in the complete semantics directly.
We report how this new assumption is proved for our instance in
\autoref{sec:proving-rscdcmd-instance}.




In order to get back to the source language our proof uses a backwards
compiler correctness assumption, again with separate compilation.
As also explained in \autoref{sec:rsc-dc}, we need to take into
account that a trace prefix $m$ in the target can be explained in the
source either by an execution producing $m$ or by one ending in an
undefined behavior (\IE producing $t {\prec} m$).
\begin{assumption}[Backward Compiler Correctness with Separate Compilation and Undefined Behavior]\label{asm:bcc}
\[
  \forall C~P~m.~(\cmp{C}\cup\cmp{P}) \doesprefix m \Rightarrow
    \exists t.~(C \cup P) {\rightsquigarrow} t \wedge (m \leq t \vee t {\prec} m)
\]
\end{assumption}

Finally, we assume that the context obtained
by back-translation can't be blamed for undefined behavior:
\ifsooner\ch{intuition missing completely}\ch{There was some above}\fi

\begin{assumption}[Blame]\label{asm:blame}
$\forall C_S:I_C.~ \forall P,P':I_P.~ \forall m~\ii{defined}.~ \forall t.$\\
If $(C_S \cup P'){\rightsquigarrow^*} m$ and
$(C_S \cup P){\rightsquigarrow} t$ and
$t \prec m$ then $m \leq t \lor t \prec_P m$.
\end{assumption}

We used Coq to prove the following theorem that puts together the
assumptions from this subsection to show \rscdcmd:

\begin{thm}\label{thm:diagram} The assumptions above imply \rscdcmd.
\end{thm}

\ifcheckpages\clearpage\fi
\section{Secure Compilation Chain}
\label{sec:compiler}
\label{sec:implementing}


\ifdesperateforspace\else
\begin{figure}
\centering
\begin{tikzpicture}[auto]
  \node (S) { Source \autoref{sec:source} };
  \node[below = of S] (I) { Compartmentalized Machine \autoref{sec:intermediate} };
  \node[below = of I, xshift=-5em] (SFI) { Bare-Metal Machine };
  \node[below = of I, xshift=5em] (MP) { Micro-Policies Machine };

  \draw[->] (S.-90) to node {compiler 
                                     } (I.90);
  \draw[->] (I.-135) to node[above,xshift=-3em] {SFI back end \autoref{sec:sfi}} (SFI.45);
  \draw[->, text width=2.5cm, yshift=-5pt] (I.-45) to node {} (MP.135);
  \node [label={[text width=2.5cm, xshift=7.2em, yshift=-8.3em]Micro-Policies back end \autoref{sec:micro-policy}}] {};
  \node[left = of SFI, xshift=3] (sT) {};
  \node (sI) at (I -| sT) {};
  \node (sS) at (S -| sT) {};
  
  \draw[rounded corners, thick, opacity=0.3]
  (sI.north east) -- (sI.north west) -- (sS.north west) node [midway, above, sloped] {Proved} -- (sS.north east);
  \draw[rounded corners, thick, opacity=0.3]
  (sT.south east) -- (sT.south west) -- (sI.south west) node [midway, above, sloped] {Tested} -- (sI.south east);
\end{tikzpicture}
\vspace{-0.8em}
\caption{Our secure compilation chain}
\label{fig:chain}
\vspace{-1em}
\end{figure}
\fi

We designed a simple proof-of-concept compilation chain to illustrate
the \rscc property.
The \iffull whole \fi compilation chain is implemented in
Coq\ifdesperateforspace\else and outlined in \autoref{fig:chain}\fi.
The \emph{source} language is a simple, unsafe imperative language
with buffers, procedures, and components (\autoref{sec:source}).
It is first compiled to an intermediate \emph{compartmentalized machine}
featuring a compartmentalized, block-structured memory,
a protected call stack, and a RISC-like instruction set augmented
with an \texttt{Alloc} instruction for dynamic storage allocation plus
cross-component \texttt{Call} and \texttt{Return} instructions (\autoref{sec:intermediate}).
We can then choose one of two back ends, which use different
techniques to enforce the abstractions of the compartmentalized machine
against realistic machine-code-level attackers,
protecting the integrity of component memories and
enforcing interfaces and cross-component call/return discipline.

When the compartmentalized machine encounters undefined behavior,
both back ends instead produce an extended trace that respects high-level
abstractions; however, they achieve this in very different ways.
The {\em SFI back end} (\autoref{sec:sfi}) targets a {\em bare-metal
  machine} that has no protection mechanisms and implements an inline
reference monitor purely in software, by instrumenting code to add address
masking operations that force each component's writes and (most) jumps
to lie within its own memory.
The {\em Micro-policies back end} (\autoref{sec:micro-policy}), on the other
hand, relies on specialized hardware~\cite{pump_asplos2015} to support a novel
tag-based reference monitor for compartmentalization.
These approaches have complementary advantages:
SFI requires no specialized hardware,
while micro-policies can be engineered to incur little
overhead~\cite{pump_asplos2015}
and are a good target for formal verification~\cite{micropolicies2015} due to their simplicity.
Together, these two back ends provide evidence that our \rscc security
criterion is compatible with any sufficiently strong
compartmentalization mechanism.
It seems likely that other mechanisms such as capability
machines~\cite{WatsonWNMACDDGL15} could also be used to implement the
compartmentalized machine and achieve \rscc.

Both back ends target variants of a simple RISC machine.
In contrast to the abstract, block-based memory model used at higher levels
of the compilation chain, the machine-level memory is a single infinite
array addressed by mathematical integers.
(Using unbounded integers is a 
simplification that we hope to remove in the future,
e.g. by applying the ideas of Mullen~\ETAL\cite{MullenZTG16}.)
%
All compartments must share this flat address space, so---without proper
protection---compromised components can access buffers out-of-bounds and
read or overwrite the code and data of other components.
Moreover, 
machine-level components can ignore the stack
discipline and jump to arbitrary locations in memory\iffull,
regardless of whether these locations are intended jump
targets or even in code memory\fi.
%

%
We establish high confidence in the security of our compilation chain
with a combination of proof and testing.
For the compiler from the source language
to the compartmentalized machine, we prove \rscc in Coq
(\autoref{sec:proving-rscdcmd-instance})
using the proof technique of \autoref{sec:proof-technique}.
For the SFI back end, we use property-based testing with
QuickChick~\cite{Paraskevopoulou15}
to systematically test \rscdcmd.\ifsooner\ch{Or even \rscc?}\fi

\subsection{Source Language}\label{sec:source}

The source language from this section was designed with simplicity in mind.
Its goal was to allow us to explore the foundational ideas of this
paper and illustrate them in the simplest possible concrete setting,
keeping our formal proofs tractable.
The language is
expression based (see \autoref{fig:source-syntax}).
A program is 
composed of an interface, a set of procedures, and a set of static buffers.
Interfaces contain the names of the procedures that the component
\emph{exports} to and \emph{imports} from other components.
Each procedure body is a single expression whose result value is
returned to the caller.
Internal and external calls share the same global, protected call stack.
Additional buffers can be allocated dynamically.
As in C, memory is manually managed; out-of-bounds accesses lead
to undefined behavior.
\iffull
Attackers can exploit these undefined behaviors to perform lower-level
attacks such as stack smashing and return-oriented programming.
\fi

Values include\ifsooner\bcp{cut (too detailed):
  mathematical\apt{+1}}\ch{isn't the undefined value interesting though?}\fi{}
integers, pointers, and
an undefined value $\top$, which is obtained when reading from an
uninitialized piece of memory or as the result of an erroneous pointer
operation. As in CompCert and LLVM~\cite{LeeKSHDMRL17}, our semantics
propagates these $\top$ values and yields an undefined behavior if a $\top$
value is ever inspected.  (The C standard, by contrast, specifies that a
program is undefined as soon as an uninitialized read or bad
pointer operation takes place.)
%



\begin{figure}
\begin{small}
  \begin{tabular}{lll}
  \texttt{e ::=} & \quad\texttt{v}               & values            \\
                 & \texttt{| local}              & local static buffer \\
                 & \texttt{| e$_1$ $\otimes$ e$_2$} & binary operations \\
                 & \texttt{| e$_1$; e$_2$}       & sequence          \\
                 & \texttt{| if e$_1$ then e$_2$ else e$_3$} & conditional \\
                 & \texttt{| alloc e}            & memory allocation \\
                 & \texttt{| !e}                 & dereferencing     \\
                 & \texttt{| e$_1$ := e$_2$}     & assignment        \\
                 & \texttt{| C.P(e)}             & procedure call    \\
                 & \texttt{| exit}               & terminate
\end{tabular}
\end{small}
\vspace{-0.8em}
\caption{Syntax of source language expressions}
\label{fig:source-syntax}
\vspace{-1em}
\end{figure}



\mypar{\iffull Block-Based \fi Memory Model}
The memory model for both source and compartmentalized machine is a
slightly simplified version of the one used in CompCert~\cite{LeroyB08}.
Each component has an infinite memory composed of finite blocks,
each an array of values.
Accordingly, a pointer is a triple $(C,b,o)$, where $C$ is the identifier of
the component that owns the block, $b$ is a unique block identifier, and $o$
is an offset inside the block.
\iflater
\bcp{We can save a little space --- and maybe avoid an argument :-)
  --- by cutting this: It's just stating what the reader should expect
  anyway.}%
For well-defined programs\bcp{is this not the case for undefined programs? I
thought {\em making} it the case was the whole point of our compilation
chain...}, pointers are unforgeable capabilities that
can point either to static buffers or
to be produced by an \texttt{alloc} operation that produces a
fresh block and that never fails.
\fi
Arithmetic operations on pointers are limited to testing equality, testing
ordering (of pointers into the same block), and changing offsets.
Pointers cannot be cast to or from integers.  Dereferencing an integer
yields undefined behavior.
For now, components are not allowed to exchange pointers; as a result,
well-defined components cannot access each others' memories at all.  We hope
to lift this restriction in the near future.
This abstract memory model is shared by the compartmentalized machine
and is mapped to a more realistic flat address space by the back ends.

\mypar{Events}
Following CompCert, we use a labeled operational semantics
whose events include all interactions of the program
with the external world (\EG system calls),
plus events tracking
\iflater
\bcp{This used to say ``interactions between components and in particular
any passing of control from one component to another.''  Are there any other
interactions between components??}
\fi
control transfers from one component to another.
Every call to an exported procedure produces a visible event
\ls$C Call P(n) C'$, recording that component \ls$C$ called procedure
\ls$P$ of component \ls$C'$, passing argument \ls$n$.
Cross-component returns are handled similarly.
All other computations, including calls and returns within the same component,
result in silent steps in the operational semantics.
%


\iflater
\ms{TODO add call/return rules and maybe local buffer}
\ch{Directly in appendix? Even there we don't have space}
\fi

\iflater
\ch{Should make it clear somewhere that technically it was a challenge
  to have a single shared stack at the source and separate stacks
  in the intermediate, but that this design makes sense.}
\fi

\subsection{The Compartmentalized Machine}\label{sec:intermediate}

\begin{figure}
\iffull
\begin{verbatim}
 r ::= R_ONE | R_COM | R_SP | R_RA | R_AUX1 | R_AUX2
\end{verbatim}
\fi

\begin{small}
\begin{tabular}{l@{\hskip0pt}l@{\hskip20pt}l}
  \texttt{instr ::=} & \quad\texttt{Nop \qquad|\qquad Halt}& \texttt{| Jal $l$}        \\
                     & \texttt{| Const $i$ -> $r$}      & \texttt{| Jump $r$}          \\
                     & \texttt{| Mov $r_s$ -> $r_d$}    & \texttt{| Call $C$ $P$}      \\
                     & \texttt{| BinOp $r_1{\otimes}r_2$ -> $r_d$} & \texttt{| Return} \\
                     & \texttt{| Load *$r_p$ -> $r_d$}  & \texttt{| Bnz $r$ $l$}       \\
                     & \texttt{| Store *$r_p$ <- $r_s$} & \texttt{| Alloc $r_1$ $r_2$} \\
  \end{tabular}
\end{small}
\vspace{-0.8em}
\caption{Instructions of compartmentalized machine}
\label{fig:intermediate-syntax}
\vspace{-1em}
\end{figure}

The compartmentalized intermediate machine aims to be
as low-level as possible while still allowing us to target our two
rather different back ends.
It features a simple RISC-like instruction set
(\autoref{fig:intermediate-syntax}) with two main abstractions: a
block-based memory model and support for cross-component calls.
The memory model 
leaves the back ends complete freedom in their layout of blocks.
The machine has a small fixed number of registers, which are the
only shared state between components.
In the syntax, $l$ represent {\em labels}, which are resolved to pointers in
the next compilation phase.

The machine uses two kinds of call stacks: a
single protected global stack for cross-component calls plus a separate
unprotected one for the internal calls of each component.
Besides the usual \texttt{Jal} and \texttt{Jump} instructions,
which are used to compile internal calls and returns, 
two special instructions, \texttt{Call} and
\texttt{Return}, are used for cross-component calls.
These are the only instructions that can manipulate
the global call stack.
\iffull
\bcp{Consider deleting this sentence:}%
The two back ends implement the abstract call
discipline in different ways using only standard RISC instructions.
\fi

%



The operational semantics rules for \texttt{Call} and \texttt{Return}
are presented in \autoref{fig:intermediate-rules}.
A state is composed of the current executing component $C$, the
protected stack $\sigma$, the memory $\ii{mem}$, the registers $\ii{reg}$ and
the program counter $\ii{pc}$.
If the instruction fetched from the program counter is a
\texttt{Call} to procedure $P$ of component $C'$, the semantics
produces an event $\alpha$ recording the caller, the callee, the
procedure and its argument, which is stored in register \texttt{R\_COM}.
The protected stack $\sigma$ is updated with a new frame containing
the next point in the code of the current component.
Registers are mostly invalidated at \texttt{Call}s; $\ii{reg}_{\top}$ has all
registers set to $\top$ and only two registers are passed on:
\texttt{R\_COM} contains the procedure's argument and \texttt{R\_RA}
contains the return address.
So no data accidentally left by the caller in the other registers
can be relied upon; instead the compiler saves and restores the registers.
Finally, there is a redundancy between the protected stack and
\texttt{R\_RA} because during the \texttt{Return} the protected frame
is used to verify that the register is used correctly; otherwise the
program has an undefined behavior.
%

\begin{figure}
 \[\frac{
    \begin{array}{l}
    \ii{fetch}(E,\ii{pc}) = \texttt{Call}~ C'~ P \qquad     C \neq C' \\
    P \in C.\ii{import} \qquad     \ii{entry}(E,C',P) = \ii{pc}' \\ 
    \ii{reg}' = \ii{reg}_{\top}[\mathtt{R\_COM} \leftarrow \ii{reg}[\mathtt{R\_COM}],
                     \mathtt{R\_RA} \leftarrow \ii{pc}+1]\\
    \alpha = C~ \mathtt{Call}(P,\ii{reg}[\mathtt{R\_COM}])~ C'
  \end{array}
  }{
    E \vdash (C,\sigma,\ii{mem},\ii{reg},\ii{pc}) \xrightarrow{\alpha} (C',(\ii{pc}+1) :: \sigma,\ii{mem},\ii{reg}', \ii{pc}')
  }
\]

\[\frac{
    \begin{array}[l]{l}
      \ii{fetch}(E,\ii{pc}) = \texttt{Return} \qquad C \neq C'\\
      \ii{reg}[\mathtt{R\_RA}] = \ii{pc}' \qquad \ii{component}(\ii{pc}') = C'\\
      \ii{reg}' = \ii{reg}_{\top}[\mathtt{R\_COM} \leftarrow \ii{reg}[\mathtt{R\_COM}]] \\
      \alpha = C~ \mathtt{Return}(\ii{reg}[\mathtt{R\_COM}])~ C'
    \end{array}
  }{
   E \vdash (C,\ii{pc}' :: \sigma,\ii{mem},\ii{reg},\ii{pc}) \xrightarrow{\alpha} (C',\sigma,\ii{mem},\ii{reg}',\ii{pc}')
  }
\]
\vspace{-0.5em}
\caption{Compartmentalized machine \iffull operational \fi semantics \iffull rules for \texttt{Call} and \texttt{Return}\fi}
\label{fig:intermediate-rules}
\vspace{-0.5em}
\end{figure}

\iflater
\ms{in the rules of the semantics stores to other components are
prevented, however in the current state we don't pass pointers, so
effectively there is no way to write (or read) to another component's
memory. Do we want to mention this as a possible future work?}
\ch{Mentioned in memory model}
\fi

\iffull
One interesting point in the semantics of this machine is the different
treatment of internal and external procedures.
For each component, the compiler introduces code to manage
a call stack for internal procedures.
In case a component exports a procedure, it can still call it
internally from another of its procedures without producing any event.
The compiler generates a single copy of each exported procedure that
can be accessed from two entry points; internal calls jump directly to
the procedure body while external calls need first to execute a
preamble which takes care of producing a \texttt{Return} once the
procedure is finished.
\fi

\subsection{\rscc Proof in Coq}
\label{sec:proving-rscdcmd-instance}
\label{sec:rscc-theorem}

We have proved that a compilation chain targeting the
compartmentalized machine satisfies \rscc, applying the technique from
\autoref{sec:proof-technique}.
As explained in \autoref{sec:rcc-informal}, the responsibility for
enforcing secure compilation can be divided among the different parts
of the compilation chain.
In this case, it is the target machine of \autoref{sec:intermediate}
that enforces compartmentalization, while the compiler itself is
simple, standard, and not particularly interesting
(so omitted here).

For showing \rscdcmd, all the assumptions from
\autoref{sec:proof-technique} are proved using simulations.
This proof was formalized in Coq, with the exception of compiler correctness
(Assumptions \ref{asm:fcc} and \ref{asm:bcc}), which is standard and essentially
orthogonal to secure compilation; eventually, we hope to scale the source
language up to a compartmentalized variant of C and reuse CompCert's mechanized
correctness proof.
Our mechanized formalization is more detailed than previous
paper proofs in the area~\cite{JuglaretHAEP16, PatrignaniC15, PatrignaniASJCP15,
  PatrignaniDP16, JeffreyR05, FournetSCDSL13, AbadiFG02, AhmedB08, AhmedB11,
  Ahmed15, NewBA16, AbadiP12, JagadeesanPRR11}.
Indeed, we are aware of only one fully mechanized proof about secure
compilation: Devriese~\ETAL's~\cite{DevriesePPK17} recent full abstraction result
for a translation from the simply typed to the untyped $\lambda$-calculus in
around 11KLOC of Coq.

Our Coq development comprises around 20KLOC, with proofs taking about 60\%.
%
Much of the code is devoted to generic models for components, traces, memory,
and undefined behavior that we expect to be useful in proofs for more complex
languages and compilers, such as CompCert.
We discuss some of the most interesting aspects of the proof below.



\paragraph{Back-translation function}\label{sec:back-mapping-function}
We proved \autoref{asm:bt} by defining a~ $\back{}$ function that takes a finite
trace prefix $m$ and a program interface $I$ and returns a whole source
program that respects $I$ and produces $m$.
Each generated component uses the local variable \texttt{local[0]} to track how
many events it has emitted.  When a procedure is invoked, it increments
\texttt{local[0]} and produces the event in $m$ whose position is given by the
counter's value.  For this back-translation to work correctly, $m$ is restricted to look
like a trace emitted by a real compiled program with an $I$ interface---in
particular, every return in the trace must match a previous call.

This back-translation is illustrated in \autoref{fig:definability} on a trace
of four events.  The generated program starts running
\texttt{MainC.mainP}, with all counters set to 0, so after testing the value
of \texttt{MainC.local[0]}, the program runs the first branch of \texttt{mainP}:
\begin{lstlisting}
local[0]++; C.p(0); MainC.mainP(0);
\end{lstlisting}
After bumping \texttt{local[0]}, \texttt{mainP} emits its first event in the
trace: the call \texttt{C.p(0)}. When that procedure starts running,
\texttt{C}'s counter is still set to 0, so it executes the first branch of
procedure \texttt{p}:
\begin{lstlisting}
local[0]++; return 1;
\end{lstlisting}
The return is \texttt{C}'s first event in the trace, and the second of the
program.  When \texttt{mainP} regains control, it calls itself recursively to
emit the other events in the trace (we can use tail recursion to
iterate in the standard way, since internal calls are silent events).
The program continues executing in this
fashion until it has emitted all events in the trace, at which point it
terminates execution.


\begin{figure*}
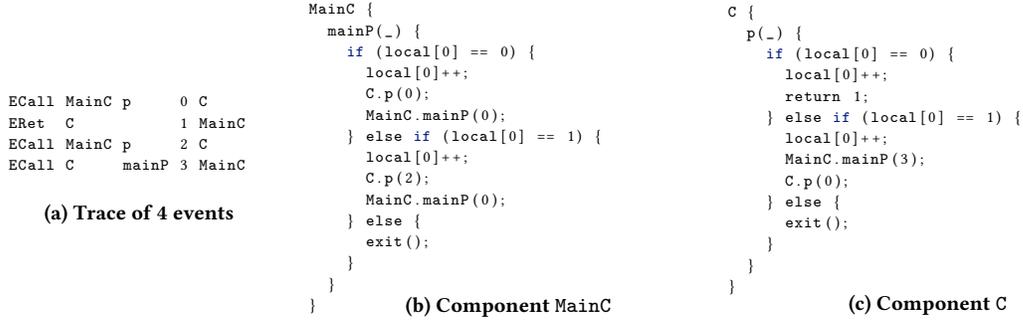


  \begin{subfigure}{0.225\textwidth}
\begin{lstlisting}
  ECall MainC p     0 C
  ERet  C           1 MainC
  ECall MainC p     2 C
  ECall C     mainP 3 MainC
\end{lstlisting}
    \caption{Trace of 4 events}
  \end{subfigure}
~
  \begin{subfigure}{0.3\textwidth}
\begin{lstlisting}
MainC {
  mainP(_) {
    if (local[0] == 0) {
      local[0]++;
      C.p(0);
      MainC.mainP(0);
    } else if (local[0] == 1) {
      local[0]++;
      C.p(2);
      MainC.mainP(0);
    } else {
      exit();
    }
  }
}
\end{lstlisting}
\vspace{-2em}
    \caption{Component \ls|MainC|}
  \end{subfigure}
~
  \begin{subfigure}{0.3\textwidth}
\begin{lstlisting}
C {
  p(_) {
    if (local[0] == 0) {
      local[0]++;
      return 1;
    } else if (local[0] == 1) {
      local[0]++;
      MainC.mainP(3);
      C.p(0);
    } else {
      exit();
    }
  }
}
\end{lstlisting}
\vspace{-1.3em}
    \caption{Component \ls|C|}
  \end{subfigure}
  \caption{Example of program with two components back-translated from
    a trace of 5 events.}
  \label{fig:definability}
\end{figure*}

\begin{thm}[Back-translation]
  The back-translation function $\uparrow$
  illustrated above satisfies \autoref{asm:bt}.
\end{thm}

\paragraph{Recomposition}
We prove \autoref{asm:recomposition} by proving a slightly more general lemma,
obtained by considering arbitrary target programs $P_T$ and $P_T'$ instead of
the compiled source programs $\cmp{P}$ and $\cmp{P'}$, and arbitrary target
context $C_T'$ instead of the compiled source context $\cmp{C_S}$.
To do so, we use a {\em three-way simulation}, stated in terms of
the underlying small-step semantics.

First, we state a {\em mergeability} relation $\sim$ between pairs of
complete states $s$ and $s'$, and prove that it is preserved
during a {\em parallel execution} guided by the trace.\ch{What
  is a ``parallel execution guided by the trace''?}
In particular, this relation ensures that the shapes of the two
stacks of the two executions are related, in that they must
capture the same interactions (calls and returns) between
program and context, despite the possibly different implementations.\ch{
  implementations of what? the procedures?}
Furthermore, we define a $\mathit{merge}$ function, implicitly parameterized
by the program and context interfaces, that defined how to merge two
mergeable states: if $s$ is a
state of $(C_T \cup P_T)$, and $s'$ is a state of $(C_T' \cup P_T')$,
then $\mathit{merge}~s~s'$ is a state of $(C_T' \cup P_T)$.
To merge two states $s$ and $s'$, one has to (1)~retain the PC and
registers of the execution that is currently in control of the run,
\IE $s$ if the program side is executing (the PC points to code belonging
to a component inside $P_T$), and $s'$ otherwise;
(2)~keep only the memory belonging to the program $P_T$ in $s$, and to the
context $C_T'$ in $s'$, and take their union;
(3)~recombine the stacks, interleaving the frames from $C_T'$ and $P_T$
in order to preserve the (compatible) structure of the original stacks.

Second, we prove the three-way simulation proper, showing that if $s_1
\xrightarrow{m} s_2$ and $s_1' \xrightarrow{m} s_2'$ where $s_1 \sim s_1'$,
then $s_2 \sim s_2'$ and $\mathit{merge}~s_1~s_1' \xrightarrow{m}
\mathit{merge}~s_2~s_2'$, where $\xrightarrow{m}$ is the reflexive transitive
closure of the step relation, producing the events in $m$.
At each step, the simulation is guided by either both programs or both contexts.
A step in one of the runs is matched by any number of silent steps in the other.
The two runs synchronize on observable events, corresponding to comparable
event-producing steps,\ch{don't see what ``comparable event-producing steps''
  is trying to add}
possibly padded with silent steps before and after.
Control alternates between programs and contexts at cross-component events.
These new definitions are more natural, completely symmetric,
and easier to work with than the previous ones about partial
semantics~\cite{AbateABEFHLPST18}.
Moreover, using the complete semantics allows us to reuse theorems already
proved for these semantics.
As a result, the resulting proofs are fully mechanized, several times smaller
than the original proofs, and several times faster to check.

While we are able to prove a generalized lemma for arbitrary target
programs and contexts in our case, we suspect that it is not possible in the
general setting.
Indeed, if the compiler plays a bigger role for enforcing security than in our
setting, one likely needs to use the fact that the programs are compiled
programs.
For example, the compiler might be inserting protections in the
compiled code at boundaries between components that ensure cross-components
calls are well-behaved.
In such a case, it is hard to see how a recomposition proof could go without
considering the compiler.
%

\begin{thm}[Recomposition]
  \autoref{asm:recomposition} is satisfied.
\end{thm}

\paragraph{Blame}
We prove \autoref{asm:blame} by noting that the behavior of the context $C_S$
can only depend on its own state and on the events emitted by the program. A bit
more formally, suppose that the states $cs_1$ and $cs_2$ have the same context
state, which, borrowing the partialization notation from above, we write as
$\texttt{par}(cs_1, I_P) = \texttt{par}(cs_2, I_P)$.  Then:
\begin{itemize}
\item If $cs_1 \xrightarrow{\alpha_1} cs_1'$,
  $cs_2 \xrightarrow{\alpha_2} cs_2'$, and $C_S$ has control in $cs_1$ and
  $cs_2$, then $\alpha_1 = \alpha_2$ and
  $\texttt{par}(cs_1', I_P) = \texttt{par}(cs_2', I_P)$.
\item If $cs_1 \xrightarrow{\tau} cs_1'$ and the program has control in $cs_1$
  and $cs_2$, then $\texttt{par}(cs_1', I_P) = \texttt{par}(cs_2, I_P)$.
\item If $cs_1 \xrightarrow{\alpha} cs_1'$, the program has control in $cs_1$
  and $cs_2$, and $\alpha \neq \tau$, then there exists $cs_2'$ such that
  $cs_2 \xrightarrow{\alpha} cs_2'$ and
  $\texttt{par}(cs_1', I_P) = \texttt{par}(cs_2', I_P)$.
\end{itemize}
By repeatedly applying these properties, we can analyze the behavior of two
parallel executions $(C_S \cup P') \doesprefix m$ and
$(C_S \cup P) \rightsquigarrow t$, with $t \prec m$.
By unfolding the definition of $t \prec m$ we get that
$\exists m' {\leq} m.~ t = m' \cdot \texttt{Undef}\left(\_\right)$.
It suffices to show that $m {\leq} t \lor t {=} m' \cdot \texttt{Undef}(P)$.
If $m = t = m' \cdot \texttt{Undef}\left(\_\right)$,
we have $m {\leq} t$, 
and we are done.
%
Otherwise, the execution
of $C_S \cup P$ ended earlier because of undefined behavior.
After producing prefix $m'$,
$C_S \cup P'$ and $C_S \cup P$ will end up in matching states $cs_1$ and $cs_2$.
Aiming for a
contradiction, suppose that undefined behavior was caused by $C_S$.  By the last
property above, we could find a matching execution step for $C_S \cup P$ that
produces the first event in $m$ that is outside of $m'$; therefore, $C_S \cup P$
cannot be stuck at $cs_2$.  Hence $t \prec_P m$.

\begin{thm}[Blame]
  \autoref{asm:blame} is satisfied.
\end{thm}

Since we have completely proved assumptions 1, 3, and 5 of \autoref{thm:diagram}
in Coq, and assumed 2 and 4, we can instantiate the generic proof technique of
\rscc from \autoref{sec:proof-technique} to obtain the final result.

\begin{thm}[RSCC]
  The compilation chain described so far in this section satisfies \rscc.
\end{thm}

\ifcheckpages\clearpage\fi
\subsection{Software Fault Isolation Back End}
\label{sec:sfi}




The SFI back end uses a special memory layout and
code instrumentation sequences to realize the desired isolation of
components in the produced program.
The target of the SFI back end is a bare-metal RISC processor
with the same instructions as the compartmentalization machine minus \texttt{Call}, \texttt{Return}, and \texttt{Alloc}. The register
file contains all the registers from the previous level, plus seven
additional registers reserved for the SFI instrumentation.

The SFI back end \iffull produces target code
satisfying \else maintains \fi the following invariants: (1) a component may
not write
outside its own data memory; (2) a component may transfer control
outside its own code memory only to entry points allowed by the
interfaces or to the return address on top of the global stack; and (3) the
global stack remains well formed.
\iflater
\bcp{Obvious:}
If any of these
invariants is violated by a program produced by the SFI back end, then
a clever attacker may exploit such vulnerabilities and cause the
\rscdcmd property to be violated.
\fi
\iflater
\nora{These invariants are necessary and sufficient for the \rscdcmd property.}
\ch{Maybe, but we haven't proved it.}
\fi

\iflater\bcp{Don't understand the next
  sentence: I thought components could not be created dynamically.}The maximum number of components
is statically determined.\fi

\iffull
We assume the target program runs
directly on the hardware.
We do not support dynamic linking or loading, nor system calls.
\fi

Figure \ref{fig:mem-layout} shows the memory layout of an application with
three components. The entire address space is divided in contiguous regions of equal size, which we will call slots. Each slot is assigned to a component or reserved for the use of the protection machinery.
Data and code are kept in disjoint memory regions and memory writes are permitted only in data regions.

An example of a logical split of a physical address is shown in Figure \ref{fig:addr-layout}. A logical address is a triple: offset in a slot, component identifier, and slot identifier unique per component.
The slot size, as well as the maximum number of the
components are constant for an application, and in Figures
\ref{fig:mem-layout} and \ref{fig:addr-layout} we have 3 components
and slots of size $2^{12}$ bits.

The SFI back end protects memory regions with instrumentation in the style
of Wahbe~\ETAL\cite{sfi_sosp1993}, but adapted to our component model.
%
Each memory update is preceded by two instructions that set the
component identifier to the current one, to prevent accidental or
malicious writes in a different component.
The instrumentation of the \texttt{Jump} instruction is similar.
The last four
bits of the offset are always zeroed and all valid targets are
sixteen-word-aligned by our back end \cite{MorrisettTTTG12}.
This mechanism, along with careful layout of instructions,
ensure that the execution of instrumentation sequences always
starts from the first instruction and continues until the end.

The global stack is implemented as a shadow stack \cite{Szekeres2013} in
memory accessible only from the SFI instrumentation sequences. Alignment of
code \cite{MorrisettTTTG12} prevents corruption of the cross-component stack
with prepared addresses and ROP attacks, since it is impossible to
bypass the instructions in the instrumentation sequence that store the
correct address in the appropriate register.

%
The \texttt{Call} instruction of the compartmentalized machine is translated to a
\texttt{Jal} (jump and link) \iffull instruction \fi followed by a
sequence of instructions that push the return address on the stack and then
restore the values of the reserved registers for the callee component.
To protect from malicious pushes that could try to use a forged
address, this sequence starts with a \texttt{Halt} at an aligned address.
%
Any indirect jump from the current component, will be aligned and will
execute the \texttt{Halt}, instead of corrupting the cross-component stack.
A call from a different component, will execute a direct jump, which
is not subject to masking operations and can thus target an unaligned
address (we check statically that it is a valid entry point).\ch{added paren}
This \texttt{Halt} and the instructions that push on the stack are
contained in the sixteen-unit block.
%


The \texttt{Return} instruction is translated to an aligned sequence:
pop from the protected stack and jump to the retrieved address. This
sequence also fits entirely in a sixteen-unit block.
%
The protection of the addresses on the stack itself is realized by the
instrumentation of all the \texttt{Store} and 
\texttt{Jump} instructions in the
program. 

\begin{figure}[t]
\centering
\includegraphics[width=1.0\linewidth]{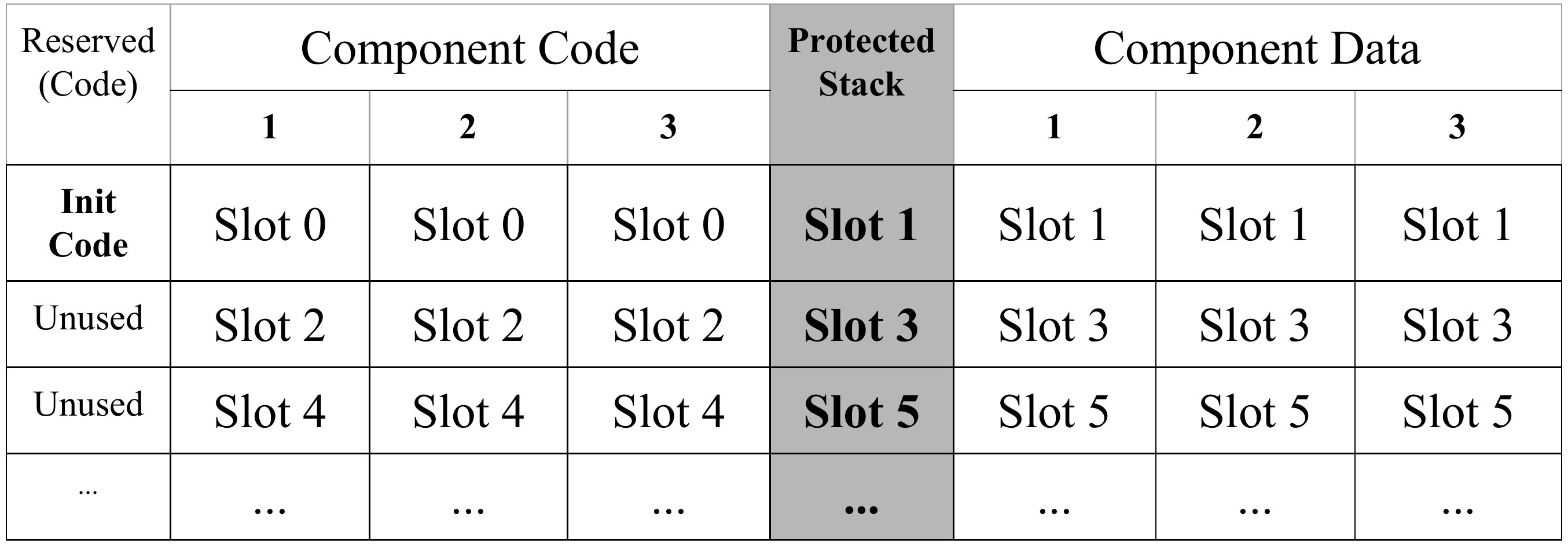}
\vspace{-1.8em}
\caption{Memory layout of three user components}
\label{fig:mem-layout}
\end{figure}
\begin{figure}[t]
\centering
\includegraphics[width=1.0\linewidth]{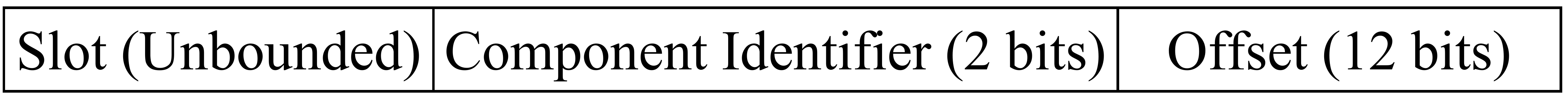}
\vspace{-1.8em}
\caption{Address Example}
\label{fig:addr-layout}
\end{figure}

\iflater
\ch{The hope is that this would become a subsection of its own
  at the end that talks about testing both back ends!}%
\fi
We used the QuickChick property-based testing tool~\cite{Paraskevopoulou15} for Coq
to test the three compartmentalization invariants described at the
beginning of the subsection. For each
invariant, we implemented a test that executes the following steps: (i)
randomly generates a \iffull syntactically \fi valid compartmentalized
machine program;
(ii) compiles it;
(iii) executes the resulting target code in a simulator and records a
property-specific trace; and (iv) analyzes the trace to verify if the
property has been violated.  We also manually
injected faults in the compiler by mutating the instrumentation
sequences of the generated output and made sure that the tests
can detect these injected errors.

More importantly, we also tested two variants of the \rscdcmd
property, which consider different parts of a whole program as the
adversarial context.
Due to the strict memory layout and the requirement that all
components are instrumented, the SFI back end cannot to link with
arbitrary target code, and has instead to compile a whole
compartmentalized machine program.
%
In a first test, we
(1) generate a whole compartmentalized machine program $P$;
(2) compile $P$;
(3) run a target interpreter to obtain trace $t_t$;
(4) if the trace is empty, discard the test;
(5) for each component $C_T$ in the trace $t_t$
(5-1) use back-translation to replace, in the program $P$, the component $C_T$ with a component $C_S$ without undefined behavior
(5-2) run the new program on the compartmentalized machine and obtain a trace $t_s$
(5-3) if the condition $t_t \leq t_s$ or $t_s {\prec}_{P \setminus \cup C_S } t_t$ is satisfied then the test passes, otherwise it fails.
Instead of performing step (5), our second test replaces in one go all
the components exhibiting undefined behavior,
obtaining a compartmentalized machine program
that should not have any undefined behavior.

\ifsooner
\ch{It would be great to extend this last test into a test for \rscc}
\fi

\iflater
\bcp{What about mutation testing for this part???}
\ch{Nora: I only had one fault injected, I eliminated the alignment. I
  didn't do the kind of experiments I had for the others.
  I think I am using the wrong email for Leo, but I hope I can work
  with him on getting the QuickChick mutation working at the Deep Spec
  Summer School when we will have the micro policy tests working too.}
\fi


\ifcheckpages\clearpage\fi
\subsection{Tag-based Reference Monitor}
\label{sec:micro-policy}

\newcommand{\xrulein}[7]{#1:&
  \{\ifthenelse{\equal{#2}{}}{}{\ii{PC}{=}#2\ifthenelse{\equal{#3}{}}{}{,}}
  \ifthenelse{\equal{#3}{}}{}{\ii{CI}{=}#3}\ifthenelse{\equal{#4}{}}{}{,}
  \ifthenelse{\equal{#4}{}}{}{\ii{NI}{=}#4}\ifthenelse{\equal{#5}{}}{}{,}
  \ifthenelse{\equal{#5}{}}{}{#5}\ifthenelse{\equal{#6}{}}{}{,}
  \ifthenelse{\equal{#6}{}}{}{#6}\ifthenelse{\equal{#7}{}}{}{,}
  \ifthenelse{\equal{#7}{}}{}{#7}
  \} &
}

\newcommand{\xruleout}[3]{
  \to\{\ii{PC}'{=}#1
  \ifthenelse{\equal{#2}{}}{}{,#2}
  \ifthenelse{\equal{#3}{}}{}{,#3}
  \}
}

Our second back end is a novel application of
a programmable tagged architecture that
allows reference monitors, called {\em micro-policies}, to be defined in
software but accelerated by hardware for performance~\cite{micropolicies2015,pump_asplos2015}.
On a micro-policy
machine, each word in memory or
registers carries a metadata tag large enough to hold a pointer to an
arbitrary data structure in memory.
As each instruction is dispatched by the processor, the opcode of the
instruction as well as the tags on the
instruction, its argument registers or memory cells, and the program counter
are all passed to a software monitor that decides whether to allow the
instruction and, if so, produces tags for the results.
The positive decisions of this monitor are cached in hardware,
so that, if another instruction is executed in the near future with
similarly tagged arguments, the hardware can allow the request immediately,
bypassing the software monitor.

This enforcement mechanism has been shown flexible enough to implement
a broad range of tag-based reference monitors, and for many of them it
has a relatively modest impact on runtime (typically under 10\%) and power
ceiling (less than 10\%), in return for some increase in energy
(typically under 60\%) and chip area (110\%)~\cite{pump_asplos2015}.%
\iflater
\ch{There was progress on energy usage, but Andre said that story is
  still work in progress and they'll have to return to it at some point}
\bcp{Let's not argue about it now, but I'm no longer convinced that (Andre
  is convinced that) these numbers are defensible...}
\fi{}
Moreover, the mechanism is simple enough so that the security of the
reference monitors can be verified formally~\cite{micropolicies2015,
  PicoCoq2013, memory-safety, ArthurPhD}.
The micro-policy machine targeted by our compartmentalizing back end builds
on a ``symbolic machine'' that Azevedo de Amorim~\ETAL
used to prove the correctness and security of several
micro-policies in Coq~\cite{micropolicies2015,
memory-safety, ArthurPhD}.
\iffull
\bcp{Next sentence could be cut:}%
This machine allows for micro-policies to be implemented at a
high level of abstraction, using Coq datatypes for the tags and Coq
functions for the behavior of the monitor.
\fi

The code generation and static linking parts of the
micro-policy back end are much simpler than for the SFI one.
The \texttt{Call} and \texttt{Return} instructions are mapped to
\texttt{Jal} and \texttt{Jump}.
The \texttt{Alloc} instruction is mapped to a monitor service that tags the
allocated memory according to the calling component.
\iffull
As in the SFI scheme,
assembly labels are resolved to concrete integer addresses,
instructions are given a simple encoding to integers, static buffers
are allocated and pointers to them also represented as integers.
\ch{all this same as for Nora, right? hoist?}
\bcp{Or just delete (even from full version)}
\nora{Yes, the labels are replaced by integers, representing addresses. I believe it's a detail that is fairly obvious and uninteresting. The alloc is implemented as a counter.}
\fi

A more interesting aspect of this back end is the way memory must be
tagged by the (static) loader based on metadata from previous compilation
stages.
Memory tags are tuples of the form
$t_m ::= (t_v, c, \ii{cs})$.
The tag $t_v$ is for the payload value.
The component identifier $c$, which we call
a color, establishes the component that owns the memory location.
Our monitor
forbids any attempt to write to memory if the color of the current
instruction is different from the color of the target location.
The set of colors $\ii{cs}$ identifies all the components that are
allowed to call to this location and is by default empty.
The value tags used by our monitor distinguish cross-component return addresses
from all other words in the system: $t_v ::= \ii{Ret}(n) ~|~ \bot$.
To enforce the cross-component stack discipline
return addresses are treated as {\em
  linear return capabilities}, \IE unique capabilities that cannot be
duplicated \cite{SAFEISA} and that can only be used to return once.
This is achieved by giving return addresses tags of the form
$\ii{Ret}(n)$, where the natural number $n$ represents the stack level
to which this capability can return.
We keep track of the current stack level using the tag of the
program counter: $t_\ii{pc} ::= \ii{Level}(n)$.
Calls increment the counter $n$, while returns decrement it.
A global invariant \iffull of the system \fi is that when the stack is at
$\ii{Level}(n)$ there is at most one capability $\ii{Ret}(m)$ for any
level $m$ from $0$ up to $n{-}1$.

\begin{figure*}[t]
\centering
\begin{center}
$
\arraycolsep=0pt
\begin{array}{rll}
\xrulein{\texttt{Nop}}%
  {t_\pc}
  {(\_,c,\_)}
  {(\_,c,\_)}
  {}
  {}
  {}
\xruleout%
  {t_\pc}
  {}
  {}
\\
\xrulein{\texttt{Const}}%
  {t_\pc}
  {(\_,c,\_)}
  {(\_,c,\_)}
  {}
  {}
  {}
\xruleout%
  {t_\pc}
  {}
  {\ii{D}{=}\bot}
\\
\xrulein{\texttt{BinOp}}%
  {t_\pc}
  {(\_,c,\_)}
  {(\_,c,\_)}
  {\ii{S}_1{=}\_}
  {\ii{S}_2{=}\_}
  {}
\xruleout%
  {t_\pc}
  {}
  {\ii{D}{=}\bot}
\\
\xrulein{\texttt{Mov}}%
  {t_\pc}
  {(\_,c,\_)}
  {(\_,c,\_)}
  {\ii{S}{=}t_v}
  {}
  {}
\xruleout%
  {t_\pc}
  {\ii{S}{=}\bot}
  {\ii{D}{=}t_v}
\\
\xrulein{\texttt{Load}}%
  {t_\pc}
  {(\_,c,\_)}
  {(\_,c,\_)}
  {}
  {M{=}(t_v,c,\_)}
  {}
\xruleout%
  {t_\pc}
  {M{=}(\bot,c,\_)}
  {D{=}t_v}
\\
\xrulein{\texttt{Load}}%
  {t_\pc}
  {(\_,c,\_)}
  {(\_,c,\_)}
  {}
  {M{=}(t_v,c',\_)}
  {}
\xruleout%
  {t_\pc}
  {M{=}(t_v,c',\_)}
  {D{=}\bot}
\\
\xrulein{\texttt{Store}}%
  {t_\pc}
  {(\_,c,\_)}
  {(\_,c,\_)}
  {}
  {M{=}(\_,c,\_)}
  {S{=}t_v}
\xruleout%
  {t_\pc}
  {M{=}(t_v,c,\_)}
  {S{=}\bot}
\\
\xrulein{\texttt{Bnz}}%
  {t_\pc}
  {(\_,c,\_)}
  {(\_,c,\_)}
  {}
  {}
  {}
\xruleout%
  {t_\pc}
  {}
  {}
\\
\xrulein{\texttt{Jal}}%
  {t_\pc}
  {(\_,c,\_)}
  {(\_,c,\_)}
  {P{=}\_}
  {}
  {}
\xruleout%
  {t_\pc}
  {}
  {\ii{RA}{=}\bot}
\\
\xrulein{\texttt{Jal}}%
  {\ii{Level}(n)}
  {(\_,c,\_)}
  {(\_,c',cs \ni c)}
  {P{=}\_}
  {}
  {}
\xruleout%
  {\ii{Level}(n{+}1)}
  {}
  {\ii{RA}{=}\ii{Ret}(n)}
\\
\xrulein{\texttt{Jump}}%
  {t_\pc}
  {(\_,c,\_)}
  {(\_,c,\_)}
  {P{=}\_}
  {}
  {}
\xruleout%
  {t_\pc}
  {}
  {\ii{RA}{=}\bot}
\\
\xrulein{\texttt{Jump}}%
  {\ii{Level}(n{+}1)}
  {(\_,c,\_)}
  {(\_,c',\_)}
  {P{=}\ii{Ret}(n)}
  {}
  {}
\xruleout%
  {\ii{Level}(n)}
  {}
  {P{=}\bot}
\end{array}
$
\end{center}


\caption{Compartmentalization micro-policy rules}
\label{fig:micro-policy}
\end{figure*}

Our tag-based reference monitor for compartmentalization is simple;
the complete definition is given in \autoref{fig:micro-policy}.
For \texttt{Mov}, \texttt{Store}, and \texttt{Load} the monitor
{\em copies} the tags together with the values, but for return addresses the
linear capability tag $\ii{Ret}(n)$ is {\em moved} from the source to
the destination.
\texttt{Load}s from other components are allowed but prevented
from stealing return capabilities.
\texttt{Store} operations are only allowed if the color of the changed
location matches the one of the currently executing instruction.
\texttt{Bnz} is restricted to the current component.
\texttt{Jal} to a different component
is only allowed if the color of the current component is
included in the allowed entry points; in this case and if we are at
some $\ii{Level}(n)$ the machine puts the return address in register
\texttt{RA} and the monitor gives it tag $\ii{Ret}(n)$ and it increments the pc
tag to $\ii{Level}(n{+}1)$.
\texttt{Jump} is allowed either to the current component or using a
$\ii{Ret}(n)$ capability, but only if we are at $\ii{Level}(n{+}1)$; in
this case the pc tag is decremented to $\ii{Level}(n)$ and the
$\ii{Ret}(n)$ capability is destroyed.
Instruction fetches are also checked to ensure that one cannot switch
components by continuing to execute past the end of a code region.
To make these checks as well as the ones for \texttt{Jal} convenient
we use the next instruction tag $\ii{NI}$ directly; in reality one can
encode these checks even without $\ii{NI}$ by using the program
counter and current instruction tags \cite{micropolicies2015}.
The bigger change compared to the micro-policy mechanism of
Azevedo de Amorim~\cite{micropolicies2015}
is our overwriting of input tags in order to
invalidate linear capabilities in the rules for \texttt{Mov},
\texttt{Load}, and \texttt{Store}.
For cases in which supporting this in hardware is not feasible we have
also devised a compartmentalization micro-policy that does not rely on
linear return capabilities but on linear entry points.\ch{It would
  be nice to explain this alternative micro-policy too!}
\ch{Another workaround might be to compile Movs, Loads, and Store
  in a very funny way, that forcefully invalidates the old capability.}


A variant of the compartmentalization micro-policy above was first
studied by Juglaret~\ETAL\cite{JuglaretHAPST15},
in an unpublished technical report.
Azevedo de Amorim~\ETAL~\cite{micropolicies2015} also devised a
micro-policy for compartmentalization, based on a rather different component model.
The biggest distinction is that our micro-policy enforces the stack
discipline on cross-component calls and returns.
%
%
We first encountered linear capabilities in the context of the SAFE
machine~\cite{SAFEISA}, although the idea seemed folklore.
Probably even closer related to our linear return capabilities are,
however, linear continuations~\cite{zdancewic02:linear_continuations, Filinski92}.

\iflater
\ch{Future work: While we have implemented this back end in Coq and
  tested on simple examples\ch{not yet true!}, formal verification of this
  back end is future work. We expect this to be much easier than for
  the SFI back end.  Not sure whether it's wise mentioning this here,
  but we do expect the property we prove for the back ends to be
  slightly different than RSCC, since the last level at which we can
  still link in code is the intermediate level. That's true for Nora,
  is that really true for Theo?}

\ch{Future work: formalizing the other micro-policy, based on
  one-use entry points.}
\fi

\iflater\ch{Cite the stack protection paper too once made public.
  What's the connection though? They do protect return addresses
  but don't enforce the stack discipline against malicious code.}\fi

\ifcheckpages\clearpage\fi
\section{Related Work}
\label{sec:related}

\ch{Is this also related to seL4 work~\cite{SewellWGMAK11, seL4:Oakland2013}?
  recent reading group on this}

\mypar{Fully Abstract Compilation,}
originally introduced in seminal work by
Abadi~\cite{Abadi99}, is phrased in terms of protecting
two partial program variants written in a {\em safe} source
language, when these are compiled and linked with a malicious target-level
context that tries to distinguish the two variants.
This original
attacker model differs substantially from the one
we consider in this paper, which protects the trace properties of
multiple mutually-distrustful components written in an
{\em unsafe} source language.

%
In this line of research,
Abadi~\cite{Abadi99} and later Kennedy~\cite{Kennedy06}
identified failures of full abstraction in the Java and C\# compilers.
%
Abadi~\ETAL~\cite{AbadiFG02} proved full abstraction of
secure channel implementations using cryptography.
Ahmed~\ETAL~\cite{AhmedB08, AhmedB11, Ahmed15, NewBA16} proved the
full abstraction of type-preserving compiler passes for functional
languages.
%
Abadi and Plotkin~\cite{AbadiP12} and
Jagadeesan~\ETAL~\cite{JagadeesanPRR11} expressed the protection
provided by address space layout randomization as a probabilistic
variant of full abstraction.
Fournet~\ETAL~\cite{FournetSCDSL13} devised a fully abstract compiler
from a subset of ML to JavaScript.
More recently,
Patrignani~\ETAL~\cite{PatrignaniASJCP15\ifdesperateforspace\else, LarmuseauPC15\fi}
studied fully abstract compilation to machine code,
starting from single modules written in simple, idealized
object-oriented and functional languages and targeting a hardware
enclave mechanism similar to Intel SGX~\cite{sgx}.
\ifsooner\bcp{Simply enumerating
  all these papers is OK (and better than nothing), but it would be even
  better to have something to say about how each of them relates to the
  present paper.}\fi

\mypar{Modular, Fully Abstract Compilation.}
Patrignani~\ETAL~\cite{PatrignaniDP16} subsequently proposed a ``modular''
extension of their compilation scheme to protecting multiple
components from each other.
The attacker model they consider is again different from ours:
\iffull instead of trying to restrict the scope of undefined behavior in the
source,\fi they focus on separate compilation of safe
languages and aim to protect linked target-level components
that are observationally equivalent to compiled components.  This
could be useful, for example, when hand-optimizing assembly produced by a
secure compiler.
%
In another thread of work, Devriese~\ETAL~\cite{DevriesePPK17} proved
modular full abstraction by approximate back-translation for a
compiler from simply typed to untyped $\lambda$-calculus.
This work also introduces a complete Coq formalization for the original
(non-modular) full abstraction proof of Devriese~\ETAL~\cite{DevriesePP16}.

\mypar{Beyond Good and Evil.}
The closest related work is that of
Juglaret~\ETAL~\cite{JuglaretHAEP16}, who also aim at protecting
mutually distrustful components written in an unsafe language.
They adapt fully abstract compilation to components, but observe
that defining observational equivalence for programs with undefined
behavior is highly problematic.
For instance, is the partial program ``{\tt int buf[5]; return buf[42]}''
\iffull observationally \fi equivalent to
``{\tt int buf[5]; return buf[43]}''?
Both encounter undefined behavior by accessing a buffer out of
bounds, so at the source level they cannot be distinguished.
However, in an unsafe language, the compiled versions of these
programs will likely read (out of bounds) different values
and behave differently.
%
Juglaret~\ETAL avoid this problem by imposing a strong limitation\iffull on
the components for which protection is guaranteed\fi: a set of
components is protected only if it cannot encounter undefined
behavior in {\em any} context.
This amounts to a {\em static} model of compromise:
all components that can possibly be compromised during
execution have to be treated as compromised from the start.
Our aim here is to show that, by moving away from full abstraction
and by restricting the temporal scope of undefined behavior, we can
support a more realistic {\em dynamic} compromise model.
As discussed below, moving away from full abstraction also makes our
secure compilation criterion easier to achieve in practice and to
prove at scale.

\mypar{Robustly Safe Compilation.}
Our criterion builds on {\em Robustly Safe Compilation} (\rsc), recently
proposed by Abate~\ETAL\cite{AbateBGHPT19}, who study several secure
compilation criteria that are similar to fully
abstract compilation, but that are phrased in terms of preserving
hyperproperties~\cite{ClarksonS10} (rather than observational equivalence)
against an adversarial context.
%
%
%
In particular, \rsc is equivalent to preservation of {\em robust
  safety}, which has been previously employed for the model checking
of open systems~\cite{KupfermanV99}, the analysis of security
protocols~\cite{GordonJ04}, and compositional
verification~\cite{SwaseyGD17}.

Though \rsc{} is a bit less extensional than fully abstract compilation
(since it is stated in terms of execution traces), it is easier to achieve.
In particular, because it focuses on safety instead of confidentiality,
the code and data of the protected program do not have to be hidden,
allowing for more efficient enforcement, \EG there is no need for fixed
padding to hide component sizes, no cleaning of registers when passing
control to the context (unless they store capabilities), and no indirection
via integer handlers to hide pointers; cross-component reads can be allowed
and can be used for passing large data.
We believe that in the future we can obtain a more practical notion
of data (but not code) confidentiality by adopting Garg~\ETAL's robust
hypersafety preservation criterion~\cite{AbateBGHPT19}.

While \rsc serves as a solid base for our work,
the challenges of protecting unsafe \iffull low-level \fi components from
each other are unique to our setting, since, like \iffull fully abstract compilation\else full abstraction\fi,
\rsc is about protecting a partial program written in a {\em safe}
source language against low-level contexts.
Our contribution is extending \rsc to reason about the dynamic
compromise of components with undefined behavior, taking advantage of
the execution traces to detect the compromise of components and to
rewind the execution along the same trace.

\mypar{Proof Techniques.}
Abate~\ETAL\cite{AbateBGHPT19} observe that, to prove \rsc, it suffices to
back-translate finite execution prefixes, and recently they propose
such a proof for a stronger criterion where multiple such executions
are involved.
In recent concurrent work, Patrignani and Garg~\cite{PatrignaniG18}
also construct such a proof for \rsc.
The main advantages of our \rscdcmd proof are that
(1) it applies to unsafe languages with undefined behavior and
(2) it directly reuses a compiler correctness result \`a la CompCert.
For safe source languages or when proof reuse is not needed our proof
could be further simplified.

Even as it stands though, our proof technique is simple
and scalable compared to previous full abstraction proofs.
While many proof techniques have been previously
investigated~\cite{AbadiFG02, AhmedB08, AhmedB11, NewBA16, AbadiP12,
  DevriesePPK17, FournetSCDSL13, JagadeesanPRR11}, fully abstract
compilation proofs are notoriously difficult, even for very simple
languages, with apparently simple conjectures surviving for decades
before being finally settled~\cite{DevriesePP18}.
The proofs of Juglaret~\ETAL~\cite{JuglaretHAEP16}
are no exception: while their compiler is similar to the one in
\autoref{sec:compiler}, their full abstraction-based proof is significantly
more complex than our \rscdcmd proof.
%
Both proofs give semantics to partial programs in terms of traces, as
was proposed by Jeffrey and Rathke~\cite{JeffreyR05} and
adapted to low-level target languages by Patrignani and
Clarke~\cite{PatrignaniC15}.
However, in our setting the partial semantics is given a one line
generic definition and is related to the complete one by
two\ifsooner\ch{standard? -- not the decomposition one}\fi{}
simulation proofs, which is simpler than proving a ``trace
semantics'' fully abstract.%
\iflater
\ch{Do we want to talk more here or in section 3.5 about the differences?}
\fi
%

\mypar{Verifying Low-Level Compartmentalization.}
Recent successes in formal verification have focused on showing correctness
of low-level compartmentalization mechanisms based on software fault
isolation~\cite{ZhaoLSR11, MorrisettTTTG12} or tagged
hardware~\cite{micropolicies2015}.
That work only considers the correctness of low-level
mechanisms in isolation, not how a secure
compilation chain makes use of these mechanisms to provide security
reasoning principles for code written in a higher-level programming
language with components.
However, more work in this direction seems underway, with
Wilke~\ETAL~\cite{WikleBBD17} working on a variant of CompCert with
SFI, based on previous work by Kroll~\ETAL~\cite{KrollSA14};
we believe \rscc or \rscdc could provide good top-level
theorems for such an SFI compiler.
In most work on verified compartmentalization~\cite{micropolicies2015,
  ZhaoLSR11, MorrisettTTTG12},
communication between low-level compartments is done by jumping to a
specified set of entry points; the model \iffull\bcp{changed ``mode'' to ``model''---OK?}\fi considered here is more
structured and enforces the correct return discipline.
Skorstengaard~\ETAL have also recently investigated a secure
stack-based calling convention for a simple capability
machine~\cite{SkorstengaardDB18}; they plan to
simplify their calling convention using a notion of linear return
capability~\cite{SkorstengaardDB18b} that seems similar \iffull in spirit \fi
to the one used in our micro-policy from \autoref{sec:micro-policy}.

\mypar{Attacker Models for Dynamic Compromise.}
While our model of dynamic compromise is specific to secure
compilation of unsafe languages, related notions of
compromise have been studied in the setting of cryptographic protocols,
where, for instance, a participant's secret keys could inadvertently be
leaked to a malicious adversary, who could then use them to impersonate the
victim~\cite{BasinC14, GordonJ05, FournetGM07, BackesGHMaffei09}.
This model is also similar to Byzantine behavior in distributed
systems~\cite{LamportSP82, CastroL02}, in which the ``Byzantine
failure'' of a node can cause it to start behaving in an arbitrary
way, including generating arbitrary data,
sending conflicting information to different parts of the system, and
pretending to be a correct node.
\iflater
\ch{If we have some extra cycles we could read \cite{BasinC14} and see
  if we find a model that's closest to ours.}
\fi


\iflater
\ch{Should maybe also relate to the recent compartmentalization work from
  PriSC: \cite{GollamudiF18}; and secure compilation too
  \cite{StrydonckDP18} -- much related to our linear capabilities,
  safe languages.}

\ch{Should see if I can find any relation to Zhong Shao's
  work~\cite{GuKRSWWZG15, RamananandroSWKF15, CostanzoSG16}}
\fi

\ifcheckpages\clearpage\fi
\section{Conclusion and Future Work}
\label{sec:conclusion}

%
We introduced \rscc, a new formal criterion for secure
compilation providing strong security guarantees despite the
dynamic compromise of components with undefined behavior.
This criterion gives a precise meaning to informal terms like {\em dynamic
  compromise} and {\em mutual distrust} 
used by
proponents of compartmentalization, and it offers a solid foundation for
reasoning about security of practical compartmentalized applications and
secure compiler chains.

\mypar{Formally Secure Compartmentalization for C.}
Looking ahead, we hope to apply \rscc to the C language by developing a
provably secure compartmentalizing compiler chain based on the CompCert
compiler.
Scaling up to the whole of C will certainly entail further
challenges such as defining a variant of C with components
and efficiently enforcing
compartmentalization all the way down.
We believe these
can be overcome by building on the solid basis built by this paper: the
\rscc formal security criterion, the scalable proof technique, and the
proof-of-concept secure compilation chain.

A very interesting extension is sharing memory between components.
Since we already allow arbitrary reads at the lowest level, it seems
appealing to also allow external reads from some of the components'
memory in the source.
The simplest would be to allow certain static buffers to be shared
with all other components, or only with some if we also extend the interfaces.
For \iffull supporting \fi
this extension the back-translation would need to set
the shared static buffers to the right values every time a
back-translated component gives up control; for this back-translation
needs to look at the read events forward in the back-translated trace
prefix.
More ambitious would be to allow pointers to dynamically allocated
memory to be passed to other components, as a form of read capabilities.
This would make pointers appear in the traces and one would need to
accommodate the fact that these pointers will vary at the different
levels in our compilation chain.
Moreover, each component produced by the back-translation would need
to record all the read capabilities it receives for later use.
Finally, to safety allow write capabilities one could combine
compartmentalization with memory safety~\cite{memory-safety, micropolicies2015}.

\mypar{Verifying Compartmentalized Applications.}
%
It would also be interesting to build verification tools based on
the source reasoning principles provided by \rscc and to use these tools to
analyze the security of practical compartmentalized applications.
\iffull
This could help further refine \rscc and provide validation of
its usefulness.
\fi
Effective verification on top of \rscc will, however, require good
ways for reasoning about the exponential number of dynamic compromise
scenarios.
One idea is to do our source reasoning with respect to a variant of
our partial semantics, which would use nondeterminism to capture the
compromise of components and their possible successive actions.
Correctly designing such a partial semantics for a complex language
is, however, challenging.
Fortunately, our \rscc criterion provides a more basic, low-TCB
definition against which to validate any fancier reasoning tools, like
partial semantics, program logics~\cite{JiaS0D15}, logical
relations~\cite{DevriesePB16}, \ETC

\iffull
Promising approaches that one could try to adapt to dynamic compromise
include Jia~\ETAL's System M~\cite{JiaS0D15}, and Devriese~\ETAL's
logical relations~\cite{DevriesePB16},
both of which allow bounding the behavior of a component based
on its interface or capabilities.

\ch{I'm starting to think that a program logic in our setting would
  better be based on existing program logics for I/0. See email
  pointers from Steven and Danel on Hennesey-Millner logic
  and how that could be mixed with Hoare logics, etc.}
\ch{Those logics seem about future traces, while we need something
  about past ones.}
\fi

\mypar{Dynamic Component Creation.}
Another interesting extension is supporting dynamic component creation.
This would make crucial use of our dynamic compromise model, since
components would no longer be statically known, and thus static
compromise would not apply, unless one severely restricts component
creation to only a special initialization phase~\cite{seL4:Oakland2013,SewellWGMAK11}.
We hope that our \rscc definition can be adapted to rewind execution
to the point at which the compromised component was created, replace
the component's code with the result of our back-translation, and then
re-execute.
This extension could allow us to also explore moving from our current
``code-based'' compartmentalization model to a ``data-based''
one~\cite{GudkaWACDLMNR15}, \EG one compartment per incoming network connection.


\mypar{Dynamic Privilege Notions.}
Our proof-of-concept compilation chain used a very simple notion of
interface to statically restrict the privileges of components.
This could, however, be extended with dynamic notions of privilege
such as capabilities and history-based access control~\cite{abadi2003access}.
In one of its simplest form, allowing pointers to be passed between
components and then used to write data, as discussed above, would
already constitute a dynamic notion of privilege, that is not captured
by the static interfaces, but nevertheless needs to be enforced to
achieve \rscc, in this case probably using some form of memory safety.

\iflater
\ch{General thought:
  It might be useful in the discussion to distinguish between static
  and dynamic privilege. Interfaces are an example of static
  privileges while capabilities are an example of dynamic ones.}
\fi

\mypar{Preserving Confidentiality and Hypersafety.}
%
It would be interesting to extend our security criterion and
enforcement mechanisms from robustly preserving safety to
confidentiality and hypersafety~\cite{AbateBGHPT19, ClarksonS10}.
For this one needs to control the flow of information at the target
level---\EG by restricting direct reads and read capabilities,
cleaning registers, \ETC
This becomes very challenging though, in a realistic attacker
model in which low-level contexts can observe time.

\ifanon\else
\mypar{Acknowledgments}
We thank the anonymous reviewers for their valuable feedback.
Yannis Juglaret first studied with us the compartmentalization
micro-policy from \autoref{sec:micro-policy}
and described it in an unpublished technical report~\cite{JuglaretHAPST15}.
J\'er\'emy Thibault helped us correctly extract the micro-policies
back end to OCaml.
We are also grateful to
Danel Ahman,
\'{E}ric Tanter,
Steven Sch\"afer, and
William Bowman,
for insightful discussions
and thoughtful feedback on earlier drafts.
This work is in part supported by
\href{https://erc.europa.eu/}{ERC} Starting Grant
\href{https://secure-compilation.github.io/}{SECOMP} (715753),
by \href{http://www.nsf.gov/awardsearch/showAward?AWD_ID=1513854}{NSF award
  Micro-Policies (1513854)},
and by DARPA grant SSITH/HOPE (FA8650-15-C-7558).
\fi

\ifanon
\clearpage
\fi
\appendix

\newcommand{\appsection}[1]{\ifieee\subsection{#1}\else\section{#1}\fi}
\newcommand{\appsubsection}[1]{\ifieee\subsubsection{#1}\else\subsection{#1}\fi}

\appsection{Class of safety properties preserved by \rscdc}
\label{sec:zp-details}



Since \rsc corresponds exactly to preserving
robust safety properties~\cite{AbateBGHPT19},
one might wonder what properties \rscdc preserves.
In fact, \rscdc corresponds exactly to preserving the following
class $Z_P$ against an adversarial context:
\begin{defn}\label{zpclosed}
$Z_P \triangleq \ii{Safety} \cap \ii{Closed}_{\prec_P}$, where
\[
\begin{array}{rl}
\ii{Safety} & \triangleq \{\pi ~|~ \forall t {\not\in} \pi.~
  \exists m {\leq} t.~ \forall t' {\geq} m.~ t' {\not\in} \pi \}\\
\ii{Closed}_{\prec_P} & \triangleq \{\pi ~|~ \forall t {\in} \pi.~
  \forall t'.~ t {\prec_P} t' \Rightarrow t' {\in} \pi \}\\
& = \{\pi ~|~ \forall t' {\not\in} \pi.~ \forall t.~ t {\prec_P} t' \Rightarrow t {\not\in} \pi \}
\end{array}
\]
\end{defn}
\newcommand{\Undef}[1]{\texttt{Undef}(#1)}
\newcommand{\main}[0]{\texttt{main}}
\newcommand{\mread}[0]{\texttt{read}}
\newcommand{\mwrite}[0]{\texttt{write}}
The class of properties $Z_P$ is defined as the intersection of
\ii{Safety} and the class $\ii{Closed}_{\prec_P}$ of properties
closed under extension of traces with undefined behavior in $P$~\cite{Leroy09}.
If a property $\pi$ is in $\ii{Closed}_{\prec_P}$ and it allows
a trace $t$ that ends with an undefined behavior in $P$---\IE
$\exists m.~ t = m\mathop{\cdot}\Undef{P}$---then $\pi$ should also allow
any extension of the trace $m$---\IE any trace $t'$ that has $m$ as a prefix.
The intuition is simple: the compilation chain is free to implement a
trace with undefined behavior in $P$ as an
arbitrary trace extension, so if the property accepts traces with
undefined behavior it should also accept their extensions.
Conversely, if a property $\pi$ in $\ii{Closed}_{\prec_P}$
rejects a trace $t'$, then for any prefix $m$ of $t'$ the property
$\pi$ should also reject the trace $m\mathop{\cdot}\Undef{P}$.
%

For a negative example that is not in $\ii{Closed}_{\prec_P}$,
consider the following formalization of the
property $S_1$ from \autoref{sec:rcc-informal}, requiring all writes
in the trace to be preceded by a corresponding read:
\[
\begin{array}{r}
S_1 {=} \{ t ~|~ \forall m~\text{\ls|d x|}.~
m\mathop{\cdot}\text{\ls|E.write(<d,x>)|} \leq t\\
\Rightarrow \exists m'.~
m'\mathop{\cdot}\text{\ls|E.read|}\mathop{\cdot}\text{\ls|Ret(x)|} \leq m \}
\end{array}
\]
While property $S_1$ is \ii{Safety} it is not $\ii{Closed}_{\prec_P}$.
Consider the trace $t' = [\text{\ls|C$_0$.main(); E.write(<d,x>)|}] \not\in S_1$
that does a write without a read and thus violates $S_1$.
For $S_1$ to be $\ii{Closed}_{\prec_P}$ it would have to reject not only $t'$,
but also [\ls|C$_0$.main(); Undef($P$)|] and \ls|Undef($P$)|, which it does not.
One can, however, define a stronger variant of $S_1$
that is in $Z_P$:
\[
\begin{array}{r}
S_1^{Z_{P}^+} {=} \{ t | \forall m~\text{\ls|d x|}.
(m\mathop{\cdot}\text{\ls|E.write(<d,x>)|} {\leq} t \vee
  m\mathop{\cdot}\text{\ls|Undef($P$)|} {\leq} t)\\
\Rightarrow \exists m'. ~m' \mathop{\cdot}
  \text{\ls|E.read|} \mathop{\cdot} \text{\ls|Ret(x)|} \leq m \}
\end{array}
\]
The property $S_1^{Z_{P}^+}$ requires any write
{\em or undefined behavior} in $P$ to be preceded by a corresponding read.
While this property is quite restrictive, it does hold (vacuously) for
the strengthened system in \autoref{fig:example-components-secure}
when taking $P = \{C_0\}$ and $C = \{C_1,C_2\}$, since we assumed that
$C_0$ has no undefined behavior.\ifsooner\ch{Less vacuous example!?}\fi

Using \iffull the class \fi $Z_P$,
we proved an equivalent \rscdc characterization:
\begin{thm}\label{zp}~\\[-1.5em]
\[
\begin{array}{rl}
\rscdc{\iff}\left(
\begin{array}{r}
\forall P~\pi {\in} Z_P.~ (\forall C_S~t.~ C_S[P] {\rightsquigarrow} t \Rightarrow t {\in} \pi )\\
\Rightarrow (\forall C_T~t.~ C_T[\comp{P}] {\rightsquigarrow} t \Rightarrow t {\in} \pi)
\end{array}
\right)
\end{array}
\]
\end{thm}
This theorem shows that \rscdc is equivalent to the preservation of
all properties in $Z_P$ for all $P$.
One might still wonder how one obtains such robust safety properties
in the source language, given that the execution traces can be influenced
not only by the partial program but also by the adversarial context.
In cases in which the trace records enough information so that one can
determine the originator of each event, 
robust safety properties can explicitly talk only about the events of
the program, not the ones of the context.
%
Moreover, once we add interfaces in \rscdcmd (\autoref{sec:rsc-dc-md}) we are
able to effectively restrict the context from directly performing
certain events (\EG certain system calls), and the robust safety
property can then be about these privileged events that the
sandboxed context cannot directly perform.
%

\ifsooner
\ch{(for Carmine) In this simple setting it would be nice to prove that
  security with dynamic compromise implies security with static
  compromise. It should be in fact be quite simple for \rscdc and
  \rscdcmd, but probably more complex for \rscc.}
\fi

\iflater\ch{Answer is yes, in fact even simpler than this}
\ch{Q for Carmine: Can $Z_P$ be written equivalently as the
  intersection of safety and
\[
Z_P' = \{\pi ~|~ t {\not\in} \pi \Rightarrow \exists m {\leq} t.~
  \forall t'.~ t' {\prec_P} m \Rightarrow t' {\not\in} \pi \}
\]
This would allow different bad prefixes for safety and for $Z_P'$ so
maybe it's not equivalent? If not, what kind of \rscdc variant would
correspond to this?}
\fi

\iflater
\ch{Q for Carmine (probably post submission, just curious): What kind
  of \rscdc variants would correspond to $Z_1$ = the intersection of
  safety and refinement closure; and to $Z_2$ = the $Z_P$ variant you
  wrote on my board where instead of $m {\leq} t$ you had $m {\leq} t
  \vee t {\prec_P} m$.  Does any of these make some sense?
  If I was to guess then $\rscdc_{Z_1}$ is just the conjunction of
  \rsc and $RC^{DC}$, so not interesting (in particular not easier to
  prove than $RC^{DC}$). On the other hand $\rscdc_{Z_2}$ can probably
  be written as follows?
\[
\begin{array}{l}
\forall P~C_T~t.~
  C_T[\cmp{P}] {\rightsquigarrow} t \Rightarrow
  \forall m.~ (m {\leq} t' \vee t' {\prec_P} m) \Rightarrow\\
  \ \ \exists C_S~t'.~ C_S[P] {\rightsquigarrow} t'
  \wedge (m {\leq} t' \vee t' {\prec_P} m)
\end{array}
\]
Does this property make sense? Is \rscdc clearly better than this?
}
\fi

One might also wonder what stronger property does one have to prove
in the source in order to obtain a certain safety property $\pi$ in
the target using an \rscdc compiler in the case in which $\pi$ is not
itself in $Z_P$.
Especially when all undefined behavior is already gone
in the target language, it seems natural to look at safety
properties such as $S_1 {\not\in} Z_P$ above that do not talk at all
about undefined behavior.
For $S_1$ above, we manually defined the stronger property
$S_1^{Z_{P}^+} {\in} Z_P$ that is preserved by an \rscdc compiler.
In fact, given any safety property $\pi$ we can easily
define $\pi^{Z_{P}^+}$ that is in $Z_P$, is stronger than $\pi$, and is
otherwise as permissive as possible:
\[
\pi^{Z_{P}^+} \triangleq \pi \cap \{t ~|~ \forall t'.~ t {\prec_P} t' \Rightarrow t' {\in} \pi \}
\]
We can also easily answer the dual question asking what is left of an
arbitrary safety property established in the source when looking at
the target of an \rscdc compiler:
\[
\pi^{Z_{P}^-} \triangleq \pi \cup \{t' ~|~ \exists t {\in} \pi.~ t {\prec_P} t' \vee t' \leq t \}
\]


\ifcheckpages
\clearpage
\fi
\ifcamera
\bibliographystyle{ACM-Reference-Format}
\else
\bibliographystyle{plainurl}
\fi
\bibliography{mp,safe}

\end{document}